\documentclass[conference]{IEEEtran}

\IEEEoverridecommandlockouts
\usepackage{multirow}
\newcommand{\IGNORE}[1]{}

\long\def\ignore#1{}
\sloppypar

\newcommand{\ApproxSign}{\raise.17ex\hbox{$\scriptstyle\sim$}}

\newcommand{\squishlist}{
   \begin{list}{$\bullet$}
    { \setlength{\itemsep}{0pt}      \setlength{\parsep}{0pt}
      \setlength{\topsep}{3pt}       \setlength{\partopsep}{0pt}
      \setlength{\listparindent}{-2pt}
      \setlength{\itemindent}{-5pt}
      \setlength{\leftmargin}{1em} \setlength{\labelwidth}{0em}
      \setlength{\labelsep}{0.5em} } }

\newcommand{\squishend}{
    \end{list}  }

\newcommand{\squishlisttwo}{
   \begin{list}{$\bullet$}
    { \setlength{\itemsep}{0pt}    \setlength{\parsep}{0pt}
      \setlength{\topsep}{0pt}     \setlength{\partopsep}{0pt}
      \setlength{\leftmargin}{2em} \setlength{\labelwidth}{1.5em}
      \setlength{\labelsep}{0.5em} } }

\usepackage{url}
\usepackage[utf8]{inputenc}
\DeclareUnicodeCharacter{2264}{$\le$}

\usepackage{booktabs}
\usepackage{cite}
\usepackage{amsmath,amssymb,amsfonts}
\usepackage{algorithmic}
\usepackage{graphicx}
\usepackage{textcomp}
\usepackage{xcolor}
\def\BibTeX{{\rm B\kern-.05em{\sc i\kern-.025em b}\kern-.08em
    T\kern-.1667em\lower.7ex\hbox{E}\kern-.125emX}}
\usepackage{makecell}
\begin{document}

\title{Enabling Software Resilience in GPGPU Applications via Partial Thread Protection}

\author{\IEEEauthorblockN{Lishan Yang}
\IEEEauthorblockA{
\textit{William \& Mary}\\
Williamsburg, VA \\
lyang11@email.wm.edu}
\and
\IEEEauthorblockN{Bin Nie}
\IEEEauthorblockA{
\textit{William \& Mary}\\
Williamsburg, VA \\
bnie@email.wm.edu}
\and
\IEEEauthorblockN{Adwait Jog}
\IEEEauthorblockA{
\textit{William \& Mary}\\
Williamsburg, VA \\
ajog@wm.edu}
\and
\IEEEauthorblockN{Evgenia Smirni}
\IEEEauthorblockA{
\textit{William \& Mary}\\
Williamsburg, VA \\
esmirni@cs.wm.edu}
}

\maketitle


\begin{abstract}
Graphics Processing Units (GPUs) are widely used by various applications in a broad variety of fields to accelerate their computation but remain susceptible to transient hardware faults (soft errors) that can easily compromise application output.
By taking advantage of a general purpose GPU application hierarchical organization in threads, warps, and cooperative thread arrays, we propose a methodology
 that identifies the resilience of threads and aims to map threads with the same resilience characteristics to the same warp. This allows engaging partial replication mechanisms for error detection/correction at the warp level. By exploring 12 benchmarks (17 kernels) from 4 benchmark suites, we illustrate that threads can be remapped into reliable or unreliable warps with only 1.63\%  introduced overhead (on average), and then enable selective protection via replication to those groups of threads that truly need it.
Furthermore, we show that thread remapping to different warps does not sacrifice application performance. 
We show how this remapping facilitates warp replication for error detection and/or correction and  achieves an average reduction  of
20.61\% and 27.15\% execution cycles, respectively comparing to standard duplication/triplication.

\end{abstract}


\section{Introduction} 
\label{sec:intro}

As general purpose GPUs (GPGPUs) are becoming increasingly susceptible to transient hardware faults (soft errors) often from cosmic radiation~\cite{DBLP:conf/dsn/FratinOLSRR18} or from operating under low voltage~\cite{killi}, their reliable operation is of critical importance.
With GPGPUs becoming omnipresent in fields such as high-performance computing (HPC), 
artificial intelligence, deep learning, virtual/augmented reality, and safety critical systems such as autonomous vehicles~\cite{eklund2013medical, pratx2011gpu, wenmei-mri,
govet-health,schmerken2009wall, compfin, gpuneural, park2008low}, transient hardware faults can lead to bit flips in storage devices including the register file and DRAM.
Such bit flips are increasing in frequency as system scales increase especially in the HPC domain~\cite{nie2016large,NieMASCOTS2017,NieDSN18}.
If bit flips occur during application execution, they may result in application crashes/hangs
or even worse in silent data corruption (SDC) where the application successfully completes execution but its output is incorrect. Executions that result in SDC outcomes are the most undesirable as they erroneously provide the user with the illusion of correct output, although cases of SDC output that is within certain user-acceptable ranges may exist~\cite{NieJS20}. 
To ensure reliable application execution, several mechanisms are widely employed including error correction codes
(ECC)~\cite{nvidia2009fermi,nvidia2014kepler,nvidia2016pascal}, but ECC cannot still provide protection to datapath errors that originate from unprotected latches in 
functional units (e.g., arithmetic logic and load-store units)~\cite{hari2015sassifi}.

Reliable execution of GPGPU applications requires 
high-overhead protection mechanisms such as check-pointing~\cite{takizawa2009checuda,laosooksathit2010lightweight} or software solutions that are based on replication.
In the GPU domain such replication can be done at different levels: at the kernel, thread, or instruction level.
At the thread level, replication is based on using redundant copies of a thread (or block of threads) and then on comparing their results~\cite{wadden2014real}. Different compiler-based implementations of this idea~\cite{wadden2014real,gupta2017compiler} aim to reduce the unavoidable synchronization overhead between the original and redundant threads.
If replication is done at the instruction level~\cite{mahmoud2018optimizing}, then the overhead of redundant multi-threading can still be significant. In addition, not all dynamic instructions are typically covered. 

In this paper, we offer an orthogonal approach that is based on the fact that  
thread resilience profiles within a GPGPU application may differ significantly -- some threads are inherently resilient, while some are not~\cite{nie2018fault}, thread resilience may also depend on application input~\cite{LishanSigmetrcis2021}.
Application resilience eventually depends on the thread organization of GPGPU application software.
In GPGPU applications threads 
are arranged at three levels: kernels, thread blocks (or cooperative thread arrays (CTAs) in CUDA terminology), and warps. Each GPU core schedules work at a granularity of warp, which is usually a group of 32 threads. Each group executes the same instruction in a lock-step manner. This is essentially the basis of single-instruction-multiple-thread (SIMT) execution.

Our thesis is that GPGPU software resilience can be achieved via {\em selective warp replication} provide that threads remapped into warps such that warps consist of threads that are either reliable or unreliable. Therefore,
if warps consist of threads that are inherently reliable, these warps (their threads or their instructions) {\em do not have to be replicated to increase their resilience}. Instead, only warps that contain unreliable threads need to be replicated. 
The advantage here comes from scheduling of warps in the single-instruction-multiple-data (SIMD) paradigm: as threads within the warps are scheduled in a lock-step way, it is a lot easier to replicate an entire warp of unreliable threads rather than replicate individual threads  within warps (or instructions within threads) and reconcile their outcome as the classic redundant multi-threading~\cite{wadden2014real,gupta2017compiler} advocates.

The process of thread remapping at the warp level is transparent to the software developer and offers a simple way to reorganize code with minimal effort.
We stress that the application resilience profile (i.e., the percentage of application executions that result in crashes/hangs, SDC, and correct executions in presence of bit flips) strongly depends on branch divergence and input data taken by different threads~\cite{DBLP:conf/dsn/LiP18}.
Since application resilience is tied to input, it cannot possibly guide software development. 
The remapping that we propose in this paper allows the developer to improve application resilience in a transparent way, either by changing the thread-warp mapping to activate replication for a fully transparent approach to the code developers, or by providing guidance to the developer to simply reorganize threads in such a manner that facilitates replication but does not interfere with the parallelization and synchronization logic of the software.

In summary, we make the following contributions:

\begin{itemize}
    \item Based on individual thread resilience we categorize the warps into three  classes: a) Reliable warps where all threads are resilient to single-bit errors, b) Unreliable warps where all threads are unreliable, and c) Mixed warps that contain both reliable and unreliable threads. We show that mixed warps are abundant in kernels.
    
    \item We propose a low-overhead partial thread protection mechanism by remapping threads such that the number of mixed warps is minimized. In other words, we change the thread to warp mapping such that distinct reliable and unreliable warp groups are formed. This facilitates the need for protecting {\em only} unreliable warps as this remapping maintains the per-thread resilience profile.

    \item We present experiments using 12 benchmarks (17 kernels) from the AxBench, CUDA, PolyBench, and Rodinia suites ~\cite{yazdanbakhsh2016axbench,cudagdb,grauer2012auto,che2009rodinia} and show that 7 of these kernels can benefit from remapping. 
    We show that remapping increases on the average the percentage of reliable warps from  23.40\% to 42.08\%, while incurring only 1.63\% execution overhead due to increased number of stalls in shared memory.

    \item By duplicating or triplicating the warps, we can easily detect when an error occurs (if duplication is used) or correct the error via triplication~\cite{wadden2014real,gupta2017compiler}.
    We show that by selectively replicating warps that contain unreliable threads after remapping (i.e., unreliable or mixed warps), we achieve average performance savings 
20.61\% and 27.15\%, for detection and protection, respectively.

\end{itemize}

The remaining of the paper is organized as follows. 
Section~\ref{sec:background} describes the background of GPU architecture and the fault model used in this paper. 
Section~\ref{sec:characterization} presents characterization regarding  various thread resilience patterns observed in the studied benchmarks.
Inspired by the characterization results, we propose a partial thread protection mechanism via remapping; the details can be found in Section~\ref{sec:remapping}.
Section~\ref{sec:evaluation} evaluates the performance gains as well as the overhead of remapping.
Then, Section~\ref{sec:related-work} discusses related work, and eventually we conclude  in Section~\ref{sec:conclusion}.

\begin{table*}[!htbp]
	\centering
\small
	\caption{Selected Benchmarks.}
	\label{table:choices_of_benchmarks}
	\begin{tabular}{cccccc}
		\toprule
        \textbf{Suite} & \textbf{Benchmark} & \textbf{Kernel Name} & \textbf{Kernel ID} & \textbf{Pct. of reliable warps} & \textbf{Pct. of reliable threads} \\ \midrule \midrule

\multirow{4}{*}{AxBench} & Jmeint & Jmeint\_kernel & K1 & 0.00\% & 55.15\% \\ \cline{2-6}
\specialrule{0em}{2pt}{2pt}
 &  Laplacian & LaplacianFilter & K1 & 49.38\%  & 54.08\%\\ \cline{2-6}
\specialrule{0em}{2pt}{2pt}
& MeanFilter & AverageFilter & K1 & 17.19\% & 26.55\%  \\ \hline
\specialrule{0em}{2pt}{2pt}
\multirow{6}{*}{CUDA} &   & executeFirstLayer & K1 & 100.00\% & 100.00\% \\  \cline{3-6}
\specialrule{0em}{2pt}{2pt}
 & NN  & executeSecondLayer & K2 & 100.00\% & 100.00\% \\  \cline{3-6}
\specialrule{0em}{2pt}{2pt}
 & (NeuralNetwork) & executeThirdLayer & K3 & 100.00\% & 100.00\% \\  \cline{3-6}
\specialrule{0em}{2pt}{2pt}
 &   & executeFourthLayer & K4 & 100.00\% & 100.00\% \\  \cline{2-6}
\specialrule{0em}{2pt}{2pt}
 & SCP & scalarProdGPU & K1 & 0.00\% & 0.00\% \\  \hline
\specialrule{0em}{2pt}{2pt}
\multirow{2}{*}{PolyBench} &  2DCONV & Convolution2D\_kernel & K1 & 0.00\% & 12.11\% \\ \cline{2-6}
\specialrule{0em}{2pt}{2pt}
 & MVT & mvt\_kernel1 & K1 & 0.00\% & 0.00\%\\ \hline	
\specialrule{0em}{2pt}{2pt}
\multirow{8}{*}{Rodinia} & \multirow{2}{*}{Gaussian} & Fan1 & K1 & 87.50\% & 90.62\% \\ \cline{3-6}
\specialrule{0em}{2pt}{2pt}

 &  & Fan2 & K2 & 63.89\% & 95.87\% \\ \cline{2-6}
\specialrule{0em}{2pt}{2pt}
 & HotSpot & calculate\_temp & K1 & 25.00\% & 43.75\% \\ \cline{2-6}
\specialrule{0em}{2pt}{2pt}
  & NearestNeighbor & euclid & K1 & 0.56\% & 0.57\% \\ \cline{2-6}
\specialrule{0em}{2pt}{2pt}
 & PathFinder & dynproc\_kernel & K1 & 8.33\% & 19.79\% \\ \cline{2-6}
 \specialrule{0em}{2pt}{2pt}

 & \multirow{2}{*}{SRAD} & reduce & K3 & 100.00\% &  100.00\% \\ \cline{3-6}
 \specialrule{0em}{2pt}{2pt}
 &  & srad & K4 & 100.00\% & 100.00\% \\


        \bottomrule

	\end{tabular}
\end{table*}

\section{Background}
\label{sec:background}
In this section,  we provide a brief introduction on the
baseline GPU architecture and the GPGPU execution model.
We also discuss the fault model, fault injection method, and application resilience profile.

\subsection{GPUs and GPGPU Application Structure}

\noindent\textbf{Baseline GPU Architecture.} A GPU typically is equipped
with a large number of cores, also known as streaming-multiprocessors (SMs)
in NVIDIA terminology~\cite{nvidia2009fermi}.
Each core has its private L1 cache, software-managed scratchpad memory, and a large register file.
An interconnection network connects all these cores to global memory, which consists of various memory channels (partitions).
Every memory channel has a shared L2 cache, and its associated memory
requests are handled by a GDDR5 memory controller.
There are various protection techniques for single-bit faults in recent commercial GPUs~\cite{nvidia2009fermi,nvidia2016pascal,nvidia2014kepler}, including single-error-correction double-error-detection (SEC-DED)
error correction codes (ECCs) that protect register files, L1/L2 caches,
shared memory and DRAM against soft errors.
Other structures such as arithmetic logic units
(ALUs), thread schedulers, instruction dispatch units, load/store units (LSUs), and interconnection
network are not protected~\cite{nvidia2009fermi,nvidia2016pascal,nvidia2014kepler}.

\begin{figure}[htbp]
	\centering
	\begin{minipage}[t]{\columnwidth}
		\centering
		\includegraphics[scale=0.3]{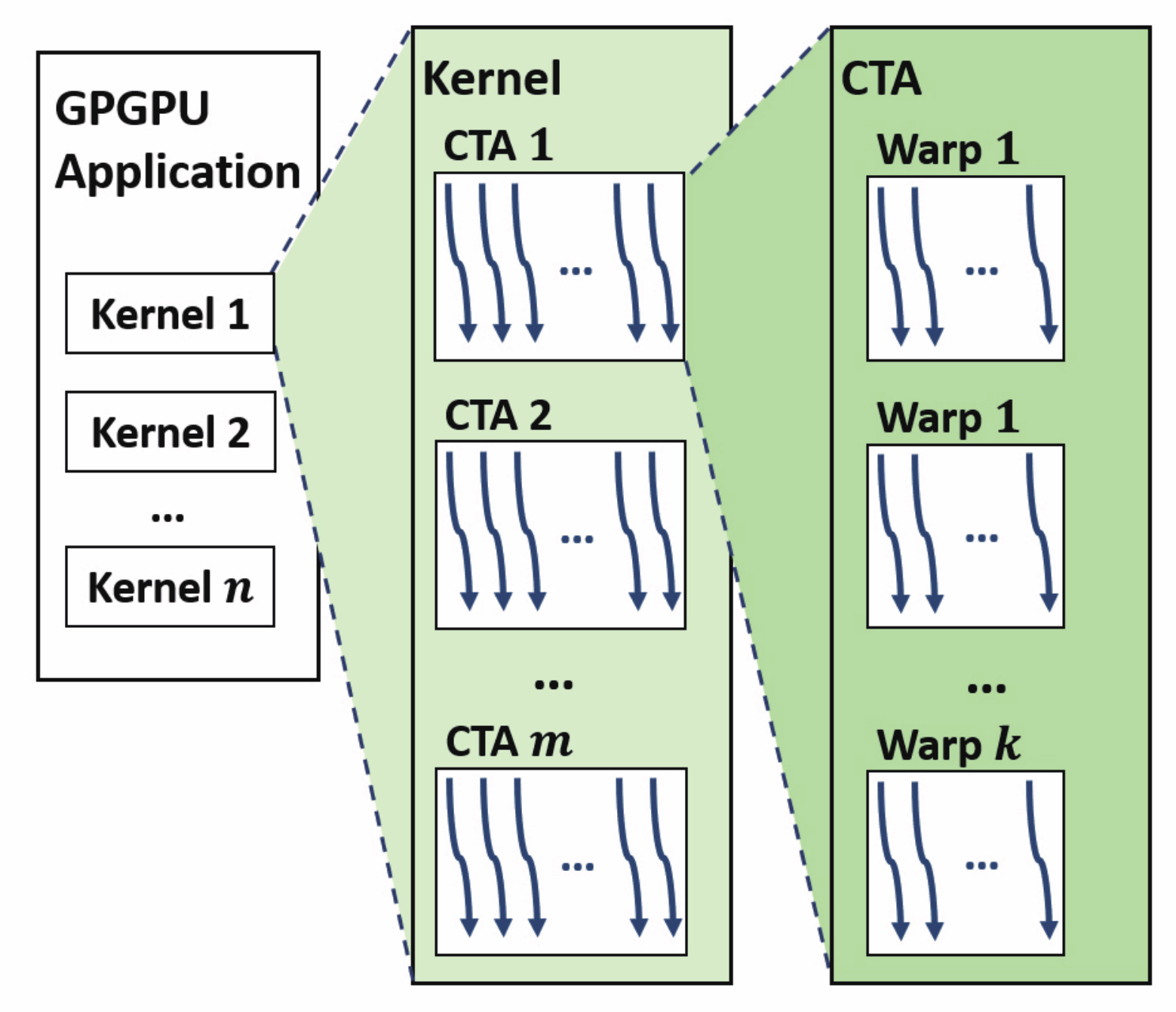}
	\end{minipage}
	\caption{GPU Software Execution Model. }
	\label{fig:gpu-arch}
\end{figure}

\vspace{2mm}
\noindent\textbf{GPGPU Software Execution Model.}
Following  the single-instruction-multiple-thread (SIMT)
philosophy~\cite{book}, GPGPU applications execute thousands of threads concurrently over large amounts of data. This helps in masking the latency and achieving high throughput.
A typical GPGPU application launches various kernels on the GPUs, see Figure~\ref{fig:gpu-arch}.
Each kernel is divided into groups of threads, known as {\it thread blocks},  which are called {\it Cooperative Thread Arrays} (CTAs) in CUDA terminology.
A CTA encapsulates all synchronization and barrier primitives among a group
of threads~\cite{book,owl-asplos13}.
 This CTA formation enables the GPU hardware to relax the execution order
 of the CTAs, for the purpose of maximizing parallelism.
Threads inside one CTA can be further divided into groups of 32 individual threads,
known as warps. 
As the most fine-grained level in terms of scheduling, warps execute a single instruction on the functional units in lock step.
This sub-division of warps is an architectural abstraction, which is transparent to the application programmer.

\subsection{Fault Model}
\label{sec:fault-model}
We assume that register files and other components such as
caches and memory are protected by ECC (which is the case in
almost all GPUs). We simulate commonly occurring computation-related
errors due to transient faults (known as soft errors) in ALUs/LSUs. These
faults can lead to wrong ALU output which would then be stored in
destination registers, or corrupted variables loaded by an LSU. This erroneous computing operation is what
we emulate by injecting faults directly to destination register
values. This is a standard experimental methodology for GPGPU
reliability studies~\cite{fang2014gpu,hari2015sassifi,llfi-gpu,nie2018fault,DBLP:conf/dsn/SangchooliePK17}.

The fault injection methodology used here closely follows the one used in \cite{tselonis2016gufi,nie2018fault}:
we flip a bit at a destination register identified by the thread id, the instruction id, and a bit position. We perform our reliability evaluations on GPGPU-Sim~\cite{gpgpu-sim} with PTXPlus mode.
GPGPU-Sim is a widely-used cycle-level GPU architectural simulator, and its PTXPlus mode provides a one-to-one mapping of instructions to actual ISA for GPUs~\cite{gpgpu-sim, tselonis2016gufi}.
Any fault injection tool or technique. (e.g., SASSIFI~\cite{hari2015sassifi} or NVBitFI~\cite{nvbitfi}) can be used for evaluating the application reliability, i.e., the technique presented in this paper does not depend on GPGPU-Sim.

\vspace{2mm}
\noindent\textbf{GPGPU Application Resilience Profile.}
For each fault injection experiment, there are three possible outcomes:
\begin{itemize}
  \item {\bf masked} output: the application output  is identical to that of fault-free execution.
  \item {\bf silent data corruption (SDC)} output: the fault injection run exits successfully without any error, but the output is incorrect.
  \item {\bf other}: the fault injection run results in a crash or hang.
\end{itemize}
To obtain the resilience profile of an application run, we conduct an  experimental campaign using the state-of-the-art fault injection methodology proposed by Nie et al.~\cite{nie2018fault} that aggressively prunes the fault space while 
achieving accuracy that is remarkably close to the ground truth.
Within the pruned fault space, we conduct one run per fault location (one single bit flip) and evaluate the application outcome as {\bf masked, SDC} or  {\bf other}.
We aggregate the outcome of all experiments 
to obtain the application {\em resilience profile}, i.e., what percentage of the runs are expected to result in masked, SDC, or other outputs.
The lower the SDC percentage, the higher the application resilience.
In this paper, we focus on reducing the  percentage of SDC outputs.
Faults that lead to masked outputs can be ignored, while faults that lead to a crash or hang are easily detected.

In this work we focus on how to improve application resilience protection when a single bit fault occurs.
The proposed methodology can be readily extended to multi-bit fault models~\cite{yang2020practical}.

\section{Characterization}
\label{sec:characterization}

We conducted experiments across 12 benchmarks (17 kernels) selected from 4 benchmark suites ~\cite{yazdanbakhsh2016axbench,cudagdb,grauer2012auto,che2009rodinia}, listed in Table~\ref{table:choices_of_benchmarks}.
The selected benchmarks cover different application domains including 3D gaming, image processing, and scientific computations.
Past work has established that different GPU threads have different resilience profiles and that the thread dynamic instruction count (iCnt) can be used as a proxy of individual thread resilience~\cite{nie2018fault,fang2014gpu}.
Indeed, fault site pruning~\cite{nie2018fault} is based on this exact concept: it demonstrates
that threads with the same dynamic instruction count have the same resilience profile, therefore it is sufficient to select one thread from each group with the same iCnt for fault injection and extrapolate the thread resilience of the entire group from a single thread. 
Our experiments further corroborate here what past work has also shown:  different threads have typically  different resilience.
Understanding the patterns of thread resilience is helpful for scheduling purposes to improve application resilience.
We categorize benchmarks into three cases: reliable, unreliable, and mixed, based on the thread resilience within each warp. We focus on warps because a warp is the smallest scheduling unit. 
In addition, it is not desirable to change the thread-CTA mapping. If done so, it breaks the synchronization and thread communication within a CTA, requiring significant effort in redesigning the parallel software logic.

The percentage of SDC outcomes across the various numbers of experiments can characterize one thread as reliable or unreliable.
For example, if the percentage of fault-injected runs that result in SDC outcomes is smaller than a small number (typically in the range from zero to 5\%, essentially if its resilience coverage is 95\%), then we characterize the thread as reliable, otherwise it is deemed unreliable.
In the following, we show some example cases.

\begin{figure}[tb]
	\centering
	\begin{minipage}[t]{\columnwidth}
		\centering
	\includegraphics[scale=0.4]{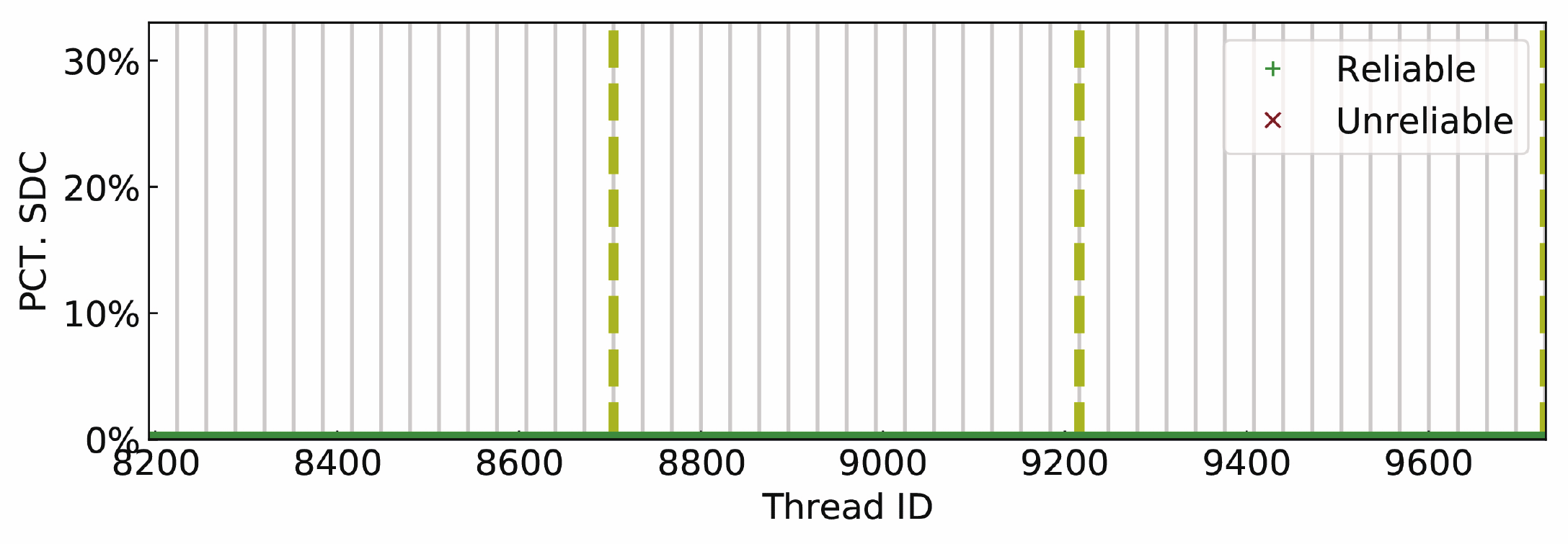}

	\end{minipage}
	\caption{All the threads are reliable in SRAD K3. Gray solid lines separate different warps, and yellow dashed lines separate different CTAs. Due to space constraint, here we only show the first 3 CTAs in the kernel. There are in total 8 CTAs in SRAD K3.}
	\label{fig:all_reliable_backprop}
	\vspace{2mm}
\end{figure}

\vspace{2mm}
{\em 1. All threads are reliable.~~~}
Some applications are very resilient to faults.
Figure~\ref{fig:all_reliable_backprop} shows the resilience scatter plot of different threads in SRAD~K3.
Threads are organized in thread launching order.
We use the gray solid lines to separate different warps, and use yellow dashed lines to separate different CTAs.
Due to space constraint, here we only show the first 3 CTAs in SRAD~K3, but the same pattern repeats across all CTAs: all threads in SRAD~K3 are reliable.
Similar to SRAD~K3,  SRAD~K4 and all NN kernels are highly resilient to soft errors.

\begin{figure}[htb]
	\centering
	\begin{minipage}[t]{\columnwidth}
		\centering
 		\includegraphics[scale=0.4]{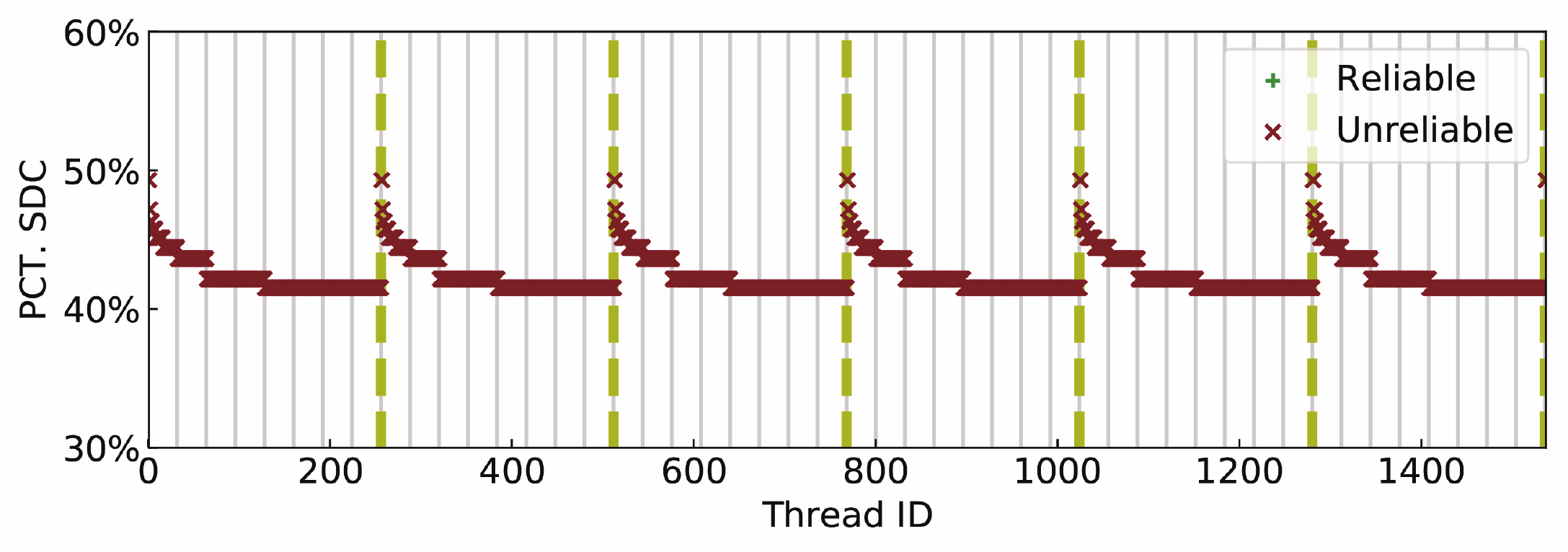}

	\end{minipage}
	\caption{All threads are unreliable in SCP. Gray solid lines separate different warps,  and yellow dashed lines separate different CTAs. There are 128 CTAs in total, but due to space constraint, we only show the first 6 CTAs.}
	\label{fig:all_unreliable_scp}
	\vspace{2mm}
\end{figure}

\vspace{2mm}
{\em 2. All threads are unreliable.~~~}
Some applications have a high probability of SDC outputs when faults are injected.
Figure~\ref{fig:all_unreliable_scp} shows the percentage of SDC outputs per thread for SCP. Here, all threads have more than 40\% SDC outputs.
An application with similar resilience behavior is MVT, with 63.82\% SDC outputs for all of its threads.

\begin{figure*}[!htb]
	\centering
 		\begin{minipage}[t]{\columnwidth}
 		\includegraphics[scale=0.4]{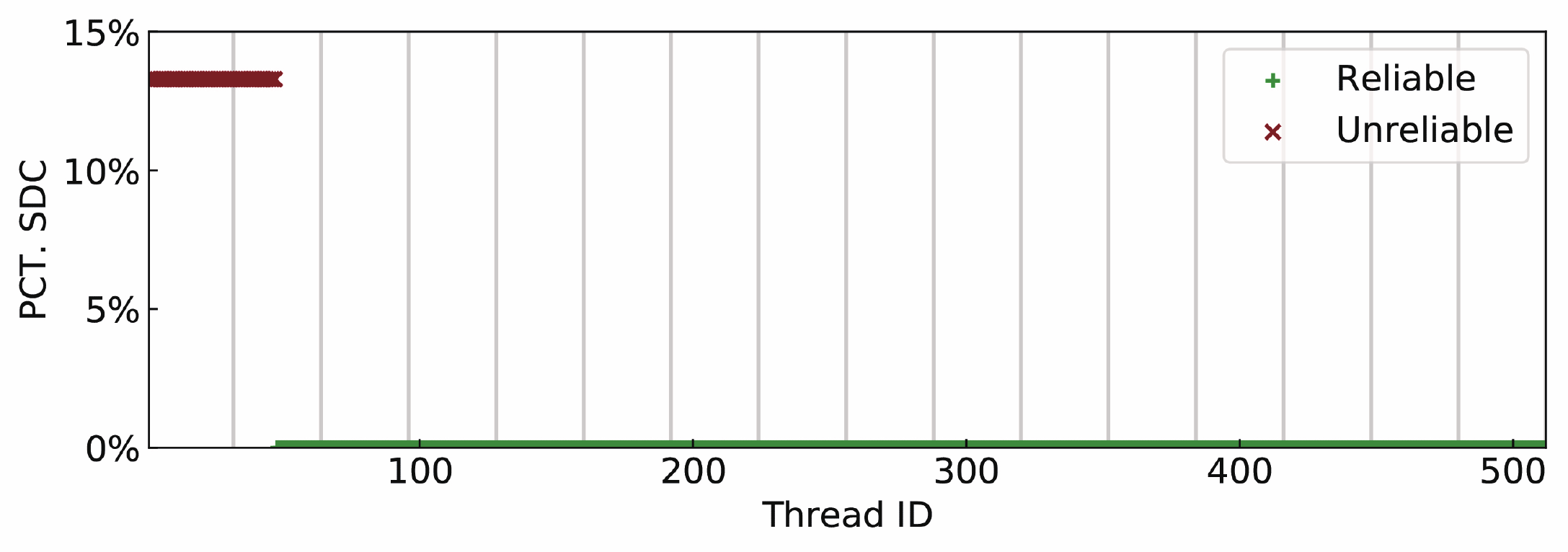}
 	\centering	
    (a) Gaussian~K1.
 	\end{minipage}
		\begin{minipage}[t]{\columnwidth}
		\centering \includegraphics[scale=0.4]{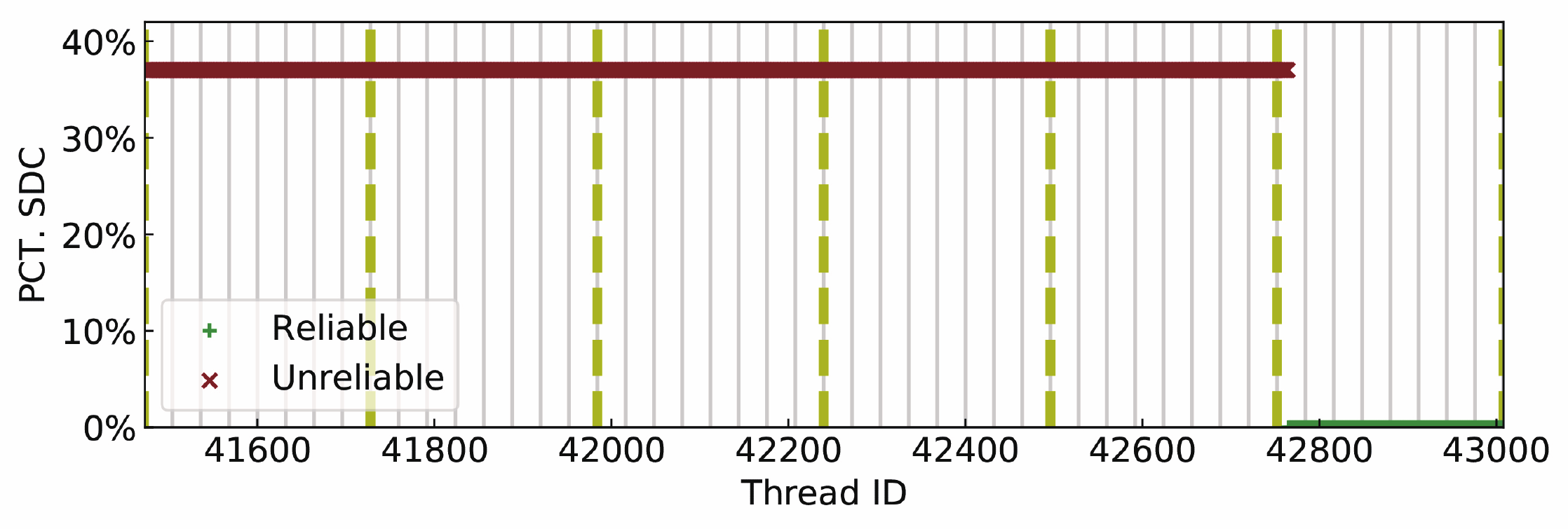}

		(b) NearestNeighbor (last 6 CTAs).
	\end{minipage}
	
	\vspace{2mm}
	\begin{minipage}[t]{\columnwidth}
		\centering \includegraphics[scale=0.4]{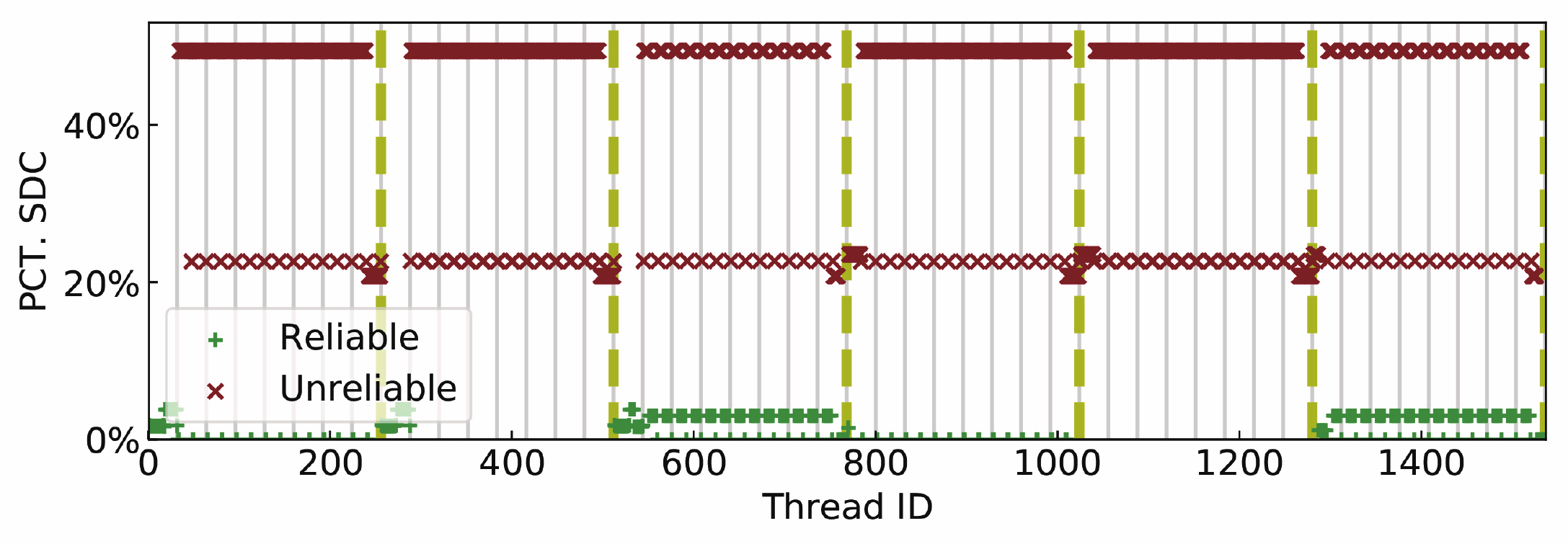}
		
    (c) HotSpot  (first 6 CTAs).
	\end{minipage}
		\begin{minipage}[t]{\columnwidth}
		\centering \includegraphics[scale=0.4]{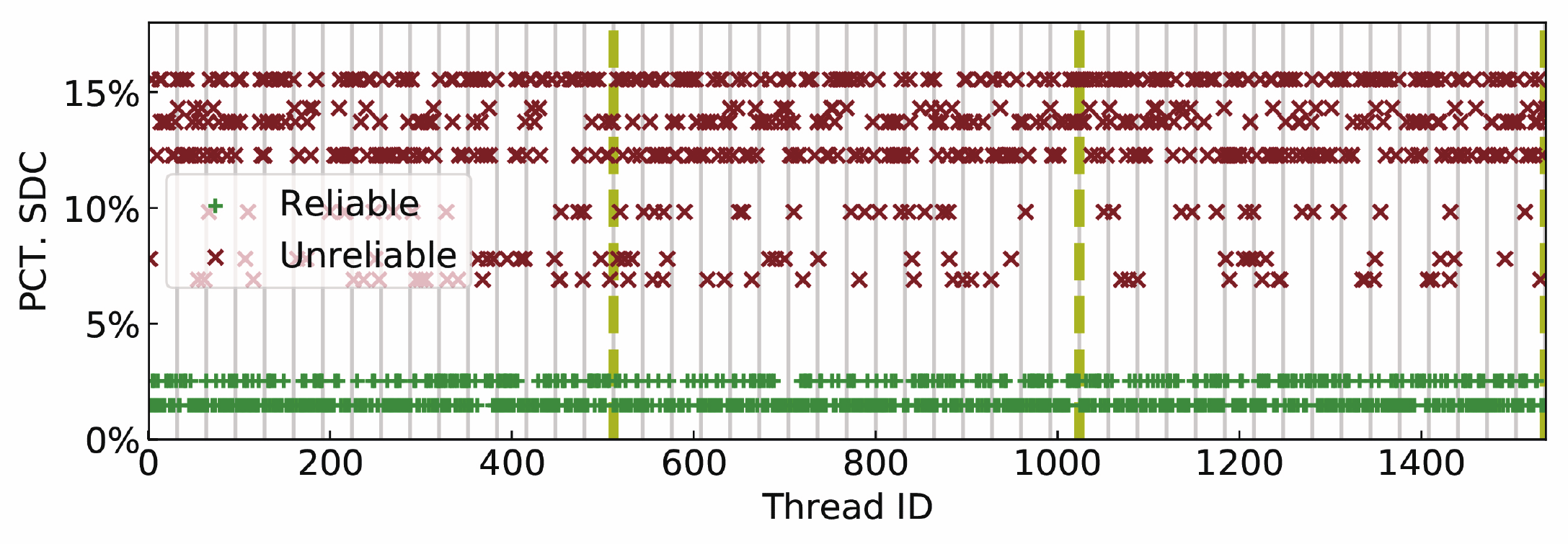}

		(d) Jmeint  (first 3 CTAs).
	\end{minipage}
\vspace{2mm}

		\begin{minipage}[t]{\columnwidth}
		\centering \includegraphics[scale=0.4]{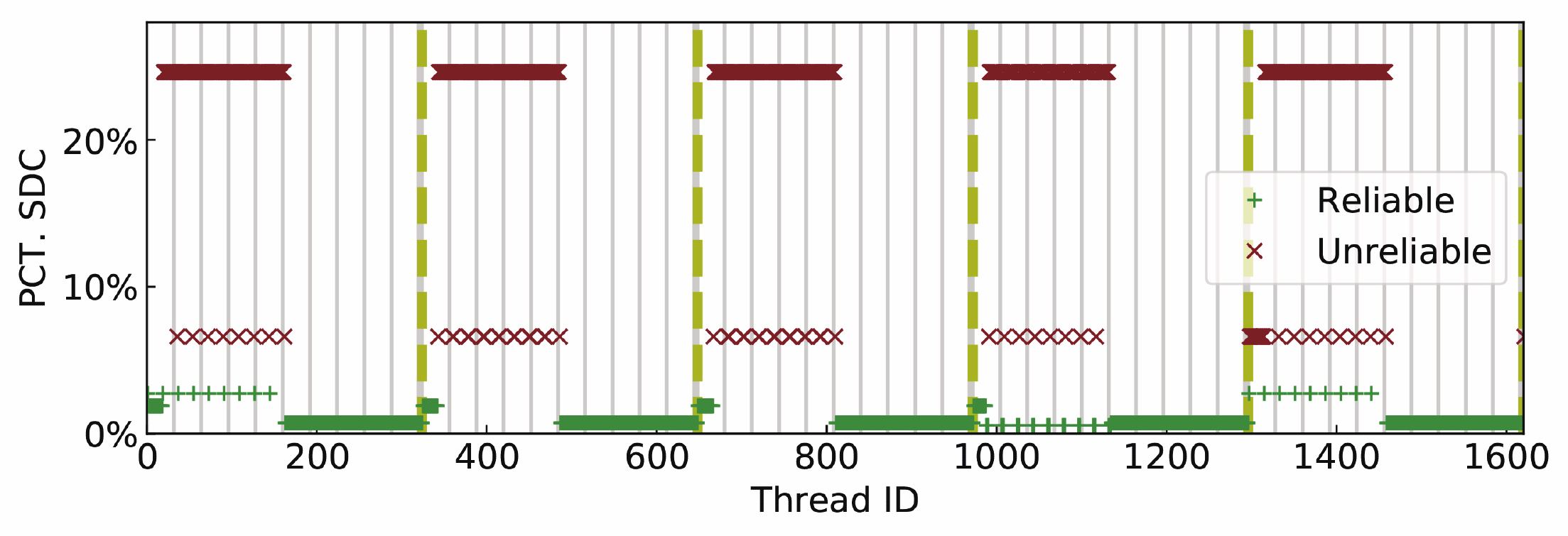}

		(e) Laplacian  (first 5 CTAs).
	\end{minipage}
		\begin{minipage}[t]{\columnwidth}
		\centering \includegraphics[scale=0.4]{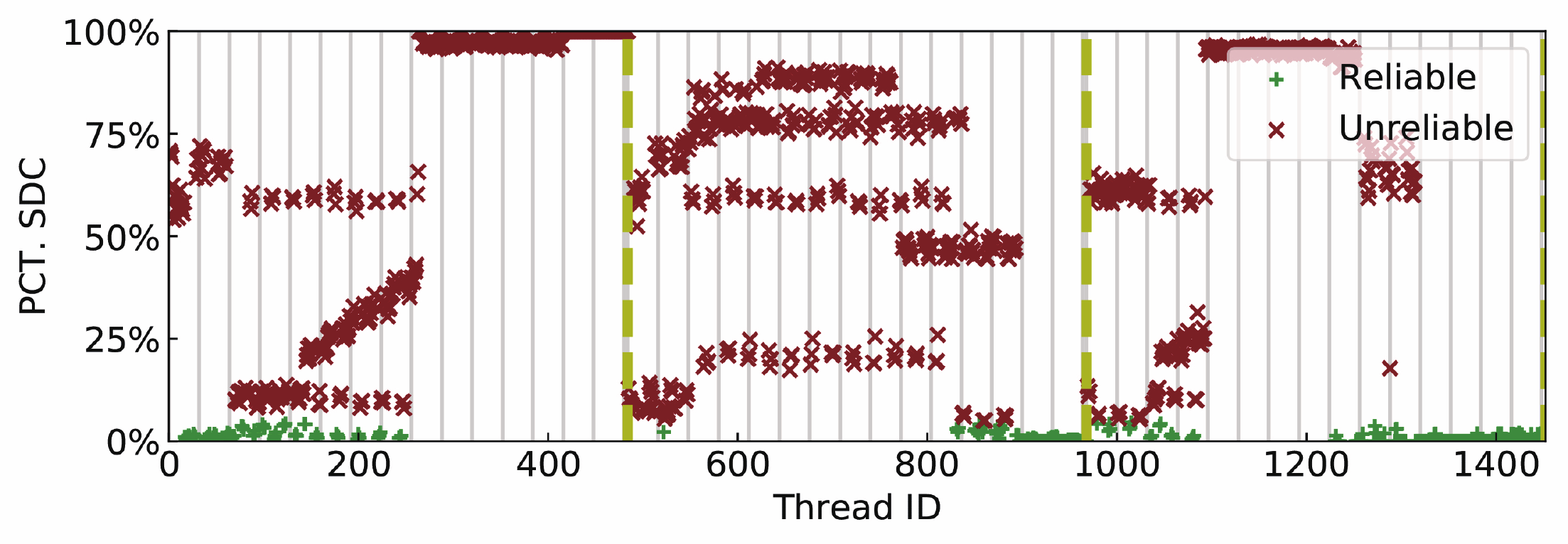}

		(f) MeanFilter  (first 3 CTAs).
	\end{minipage}
	
	\vspace{2mm}
		\begin{minipage}[t]{\columnwidth}
		\centering \includegraphics[scale=0.4]{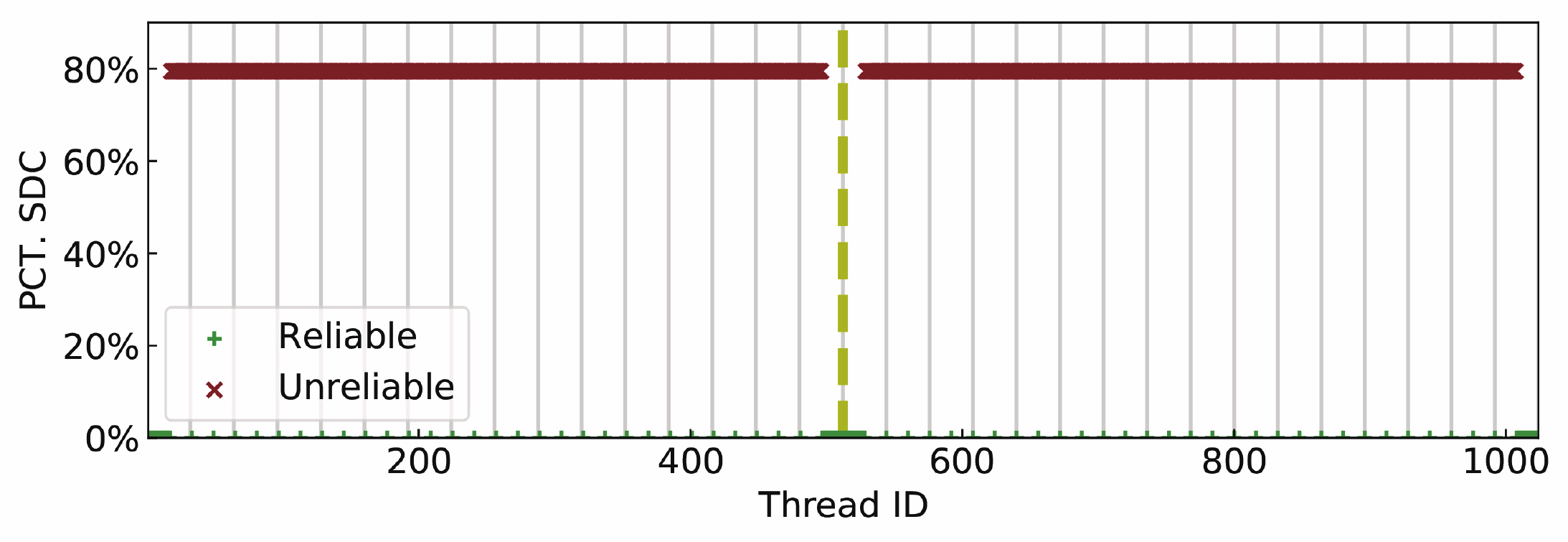}

		(g) 2DCONV.
	\end{minipage}
			\begin{minipage}[t]{\columnwidth}
		\centering \includegraphics[scale=0.4]{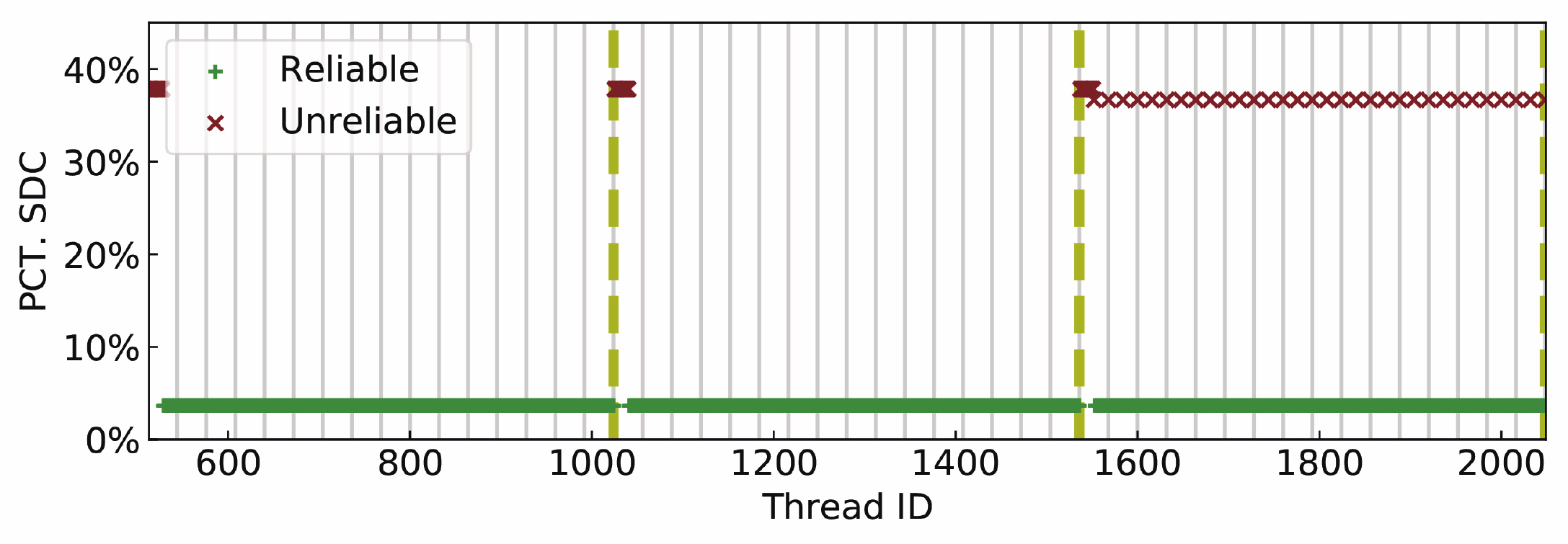}

		(h) Gaussian~K2 (first 3 CTAs).
	\end{minipage}
	\vspace{2mm}
	
	\begin{minipage}[t]{\columnwidth}
		\centering \includegraphics[scale=0.4]{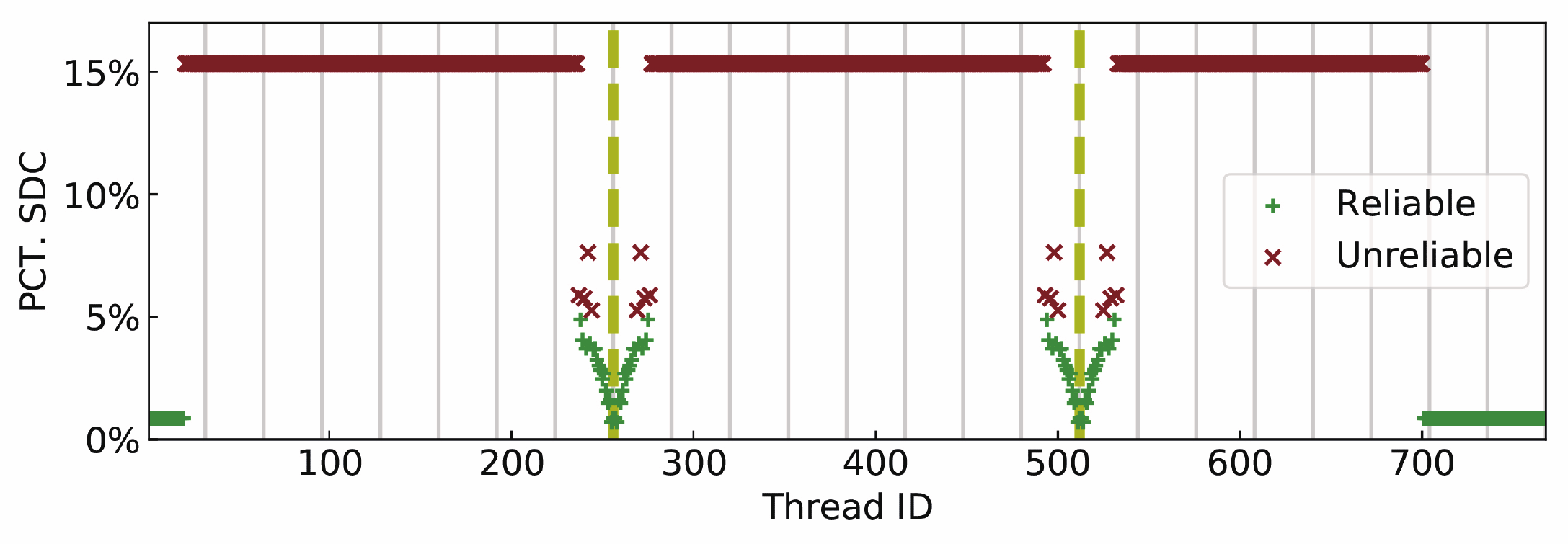}

		(i) PathFinder.
	\end{minipage}
	\caption{Reliable and unreliable threads exist together in the same warp. Due to space constraints, we only show the a part of the CTAs for NearestNeighbor, HotSpot, Jmeint, Laplacian, MeanFilter, and Gaussian~K2.}
	\label{fig:reliable_unreliable_together}
\end{figure*}

\vspace{2mm}
{\em 3. Mixed Reliable and Unreliable Threads within Warps.~~~}

Reliable and unreliable threads can co-exist in the same warp (and consequently CTA), see Figure~\ref{fig:reliable_unreliable_together}.
Reliable threads are marked with a green `+', while red `x' represents unreliable threads and marks their SDC probability.
We start from two simple benchmarks: Gaussian~K1 and NearestNeighbor (Figure~\ref{fig:reliable_unreliable_together}(a) and (b), respectively).
For Gaussian~K1, there are in total 512 threads organized in one CTA only.
The first 48 threads are unreliable, and the remaining threads are very resilient (their  percentage of SDC outputs is 0\%).
NearestNeighbor shows a similar resilience pattern: threads at the beginning are unreliable, those that are launched later are reliable.
There are in total 168 CTAs in NearestNeighbor. Due to the space constraint, here we only show the last 6 CTAs which can best express the idea of well-organized warps.
For both Gaussian~K1 and  NearestNeighbor, reliable threads and unreliable threads are already organized separately within different warps (with the exception of a single warp either at the start for Gaussian or at the tail for NearestNeighbor that contains both reliable and unreliable threads).

However, there are  benchmarks where their threads are not that well-organized.
For HotSpot, shown in Figure~\ref{fig:reliable_unreliable_together}(c), reliable  and unreliable threads are mixed within different warps.
Due to space constraint, here we only show the first 6 CTAs at the beginning for HotSpot.
As shown in Table~\ref{table:choices_of_benchmarks},
the percentage of reliable warps (warps containing only reliable threads) is 25\% for HotSpot,
but there are in total 43.75\% reliable threads.

Similarly, in Jmeint, there is no reliable warp, because all of the warps have both reliable and unreliable threads, as shown in Figure~\ref{fig:reliable_unreliable_together}(c).
For Jmeint more than half (55.15\%) of the threads are reliable.
However, since they are mixed in CTAs with the remaining 44.85\% unreliable threads, 
protecting via replication would require replication of the entire kernel, i.e.,  every warp.
Similar observations apply to Laplacian, MeanFilter, 2DCONV, Gaussian~K2, and PathFinder, see~Figure~\ref{fig:reliable_unreliable_together}(e)-(h).

Figure~\ref{fig:reliable_unreliable_together} clearly illustrates that there is ample scope for partial protection:
If we group threads judiciously, then we can increase the percentage of reliable warps, and avoid redundant protection of warps (threads) that are anyway resilient.
Table~\ref{table:benchmark-category} summarizes the benchmark categorization.

\begin{table}[htbp]
	\centering
       \caption{{Benchmark categories}}
       \label{table:benchmark-category}

     \small
     \begin{tabular}{p{0.9cm} c l}
    \toprule
\multicolumn{2}{c}{\textbf{Category}}  & \textbf{Benchmark} \\
\midrule \midrule
\multicolumn{2}{c}{\multirow{2}{*}{All threads are reliable}}  & NN~K1, NN~K2, NN~K3, \\
\specialrule{0em}{1pt}{1pt} 
& & NN~K4, SRAD~K3, SRAD~K4\\ \hline
\specialrule{0em}{1pt}{1pt}
\multicolumn{2}{c}{All threads are unreliable} & SCP, MVT \\ \hline

\specialrule{0em}{1pt}{1pt}
\multirow{5}{*}{\makecell{Mixed \\ warps}} & Well-organized & Gaussian~K1, NearestNeighbor \\ \cline{2-3}
\specialrule{0em}{1pt}{1pt}
& \multirow{3}{*}{Need Remapping} & Jmeint, MeanFilter, 2DCONV, \\
\specialrule{0em}{1pt}{1pt}
& & HotSpot, Gaussian~K2,   \\
\specialrule{0em}{1pt}{1pt}
& & PathFinder, Laplacian \\
 \bottomrule
     \end{tabular}
\end{table}

\vspace{2mm}
{\em Summary.~~~}
From the reliability perspective, there is no need to protect reliable threads.
Benchmarks where all threads are reliable, result in reliable kernel executions.
Similarly, for benchmarks that have warps that are all unreliable, protection needs to be applied to the entire kernel.
Approaching kernel reliability from
the scheduling perspective, threads are grouped and scheduled in units of warps, which is transparent to software developers.
For benchmarks that consist of both reliable and unreliable threads, we explore ways to {\it remap} threads into warps such that warps consist of {\em only} reliable or unreliable threads.
If this is done, then it is not necessary to protect the application fully but instead focus on protecting unreliable warps only.

\begin{figure}[tbp]
	\centering
	\begin{minipage}[t]{\columnwidth}
		\centering
		\includegraphics[scale=0.5]{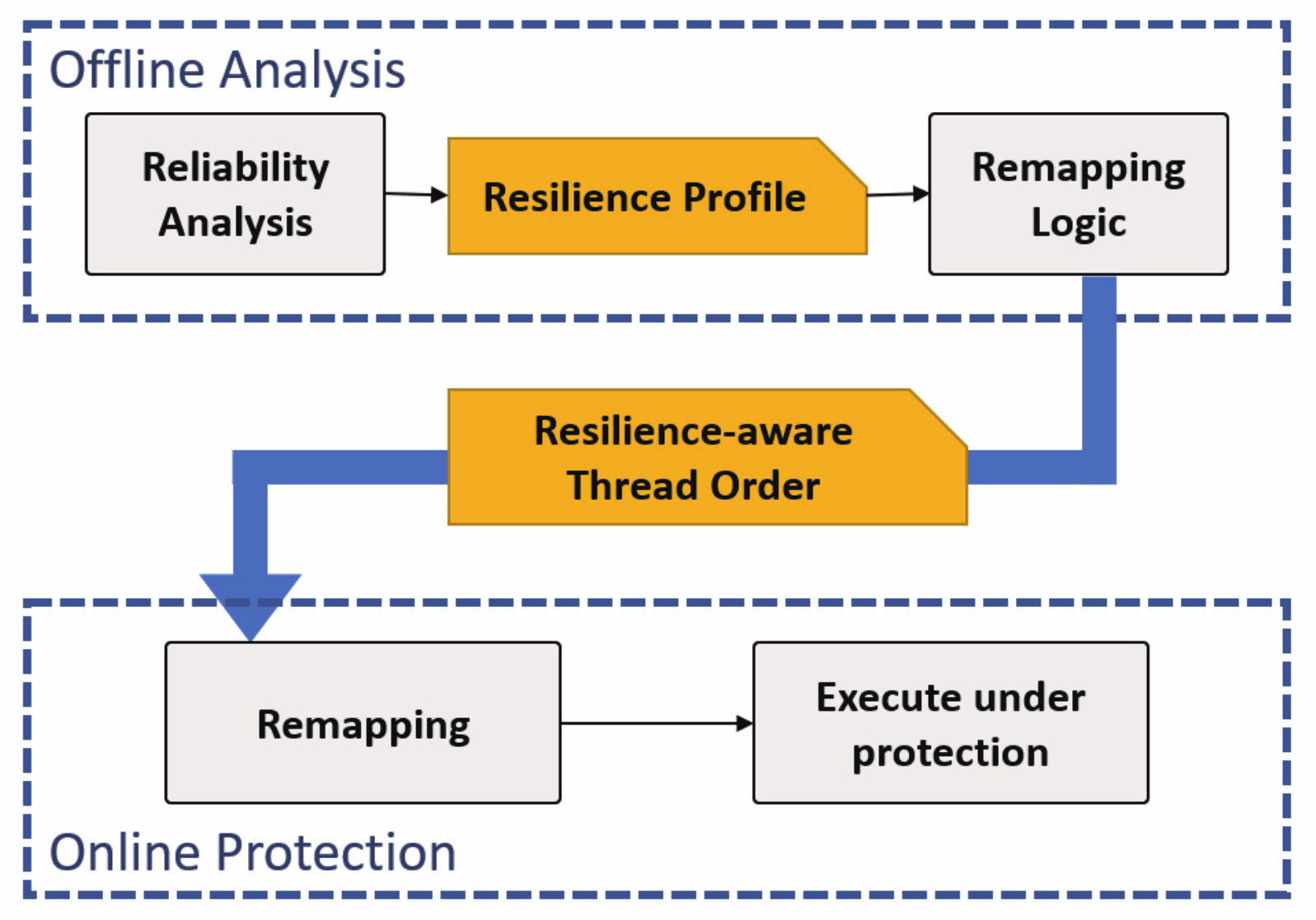}
	\end{minipage}
	\caption{Workflow.}
	\label{fig:workflow}
\end{figure}

\section{Resilient Software Protection via Remapping}
\label{sec:remapping}

In~\cite{nie2018fault}, threads are identified as the most important GPGPU component, and the resilience pattern of an application can be derived from thread resilience.
Here, we propose a low-overhead partial protection mechanism  that leverages thread resilience patterns via remapping.
The main idea is to remap threads to warps for the purpose of separating reliable and unreliable threads, as scheduling of threads can be done at the warp granularity.
By addressing the problem at the warp level, we propose to recompute warps that contain unreliable  threads,  essentially offering {\em partial protection} to a subset of warps and not the entire kernel, without compromising application reliability.

Figure~\ref{fig:workflow} shows the workflow of the proposed protection mechanism. There are two components: 1) Offline analysis, to obtain the resilience-aware thread order, and 2) Online protection, which uses this resilience-aware thread order to achieve low-overhead protection.
\begin{enumerate}
\vspace{1mm}
    \item {\bf Offline Analysis. }
        For any target kernel and for a specific input, the resilience of each thread needs to be first obtained.
        There is no restriction on which method of reliability analysis is used.
        Fault injection campaigns~\cite{hari2015sassifi,fang2014gpu,nie2018fault}, ACE (Architecturally Correct Execution) analysis, 
        or a combined method leveraging both fault injection and ACE analysis~\cite{vallero2019combining} can be used. 
        The only requirement is that  the resilience of every thread needs to be evaluated.
        This is not difficult to do, despite the fact that most GPGPU applications have tens of thousands of threads, because for most benchmarks threads with the same DI count have the same resilience behavior~\cite{fang2014gpu,nie2018fault}.
        This reduces the number of experiments that need to be done to obtain the thread resilience profile.
        In this work, we evaluate thread and kernel resilience using the fault site pruning technique~\cite{nie2018fault}.

        We first identify the resilience profile of threads in their launching order (see Figure~\ref{fig:reliable_unreliable_together}). 
        Then, threads are re-ordered into warps following the remapping logic: threads with similar resilience are remapped into the same warp. Detailed explanation is in Subsection~\ref{sec:remapping-logic}.
        
        \vspace{1mm}
        
    \item {\bf Online Protection.}
    Based on the resilience-aware thread order obtained from offline analysis, we can remap threads before execution. 
    The actual remapping idea can be implemented in various ways. In this work, we 
    directly change the thread-warp mapping (see Subsection~\ref{sec:remapping-logic} for implementation details). 
    After remapping, threads are executed, and error detection/correction is applied to unreliable warps only.
    Error detection/correction can be implemented and applied in various ways. Here, we use warp duplication for error detection, and warp triplication for error correction. Details are given in Subsection~\ref{sec:duplication}.

\end{enumerate}

\begin{figure}[tbp]
	\centering
	\begin{minipage}[t]{\columnwidth}
		\centering
		\includegraphics[scale=0.38]{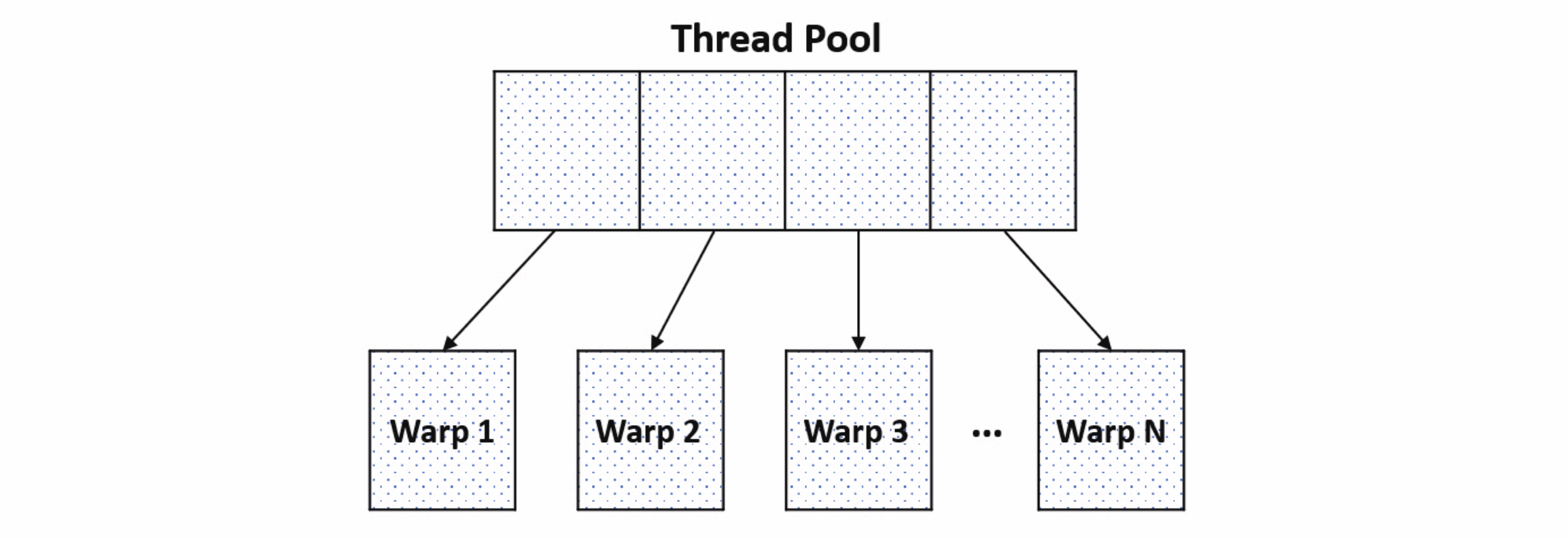}

		(a) Thread-warp mapping.
	\end{minipage}
\vspace{\baselineskip}

	\begin{minipage}[t]{\columnwidth}
		\centering
		\includegraphics[scale=0.38]{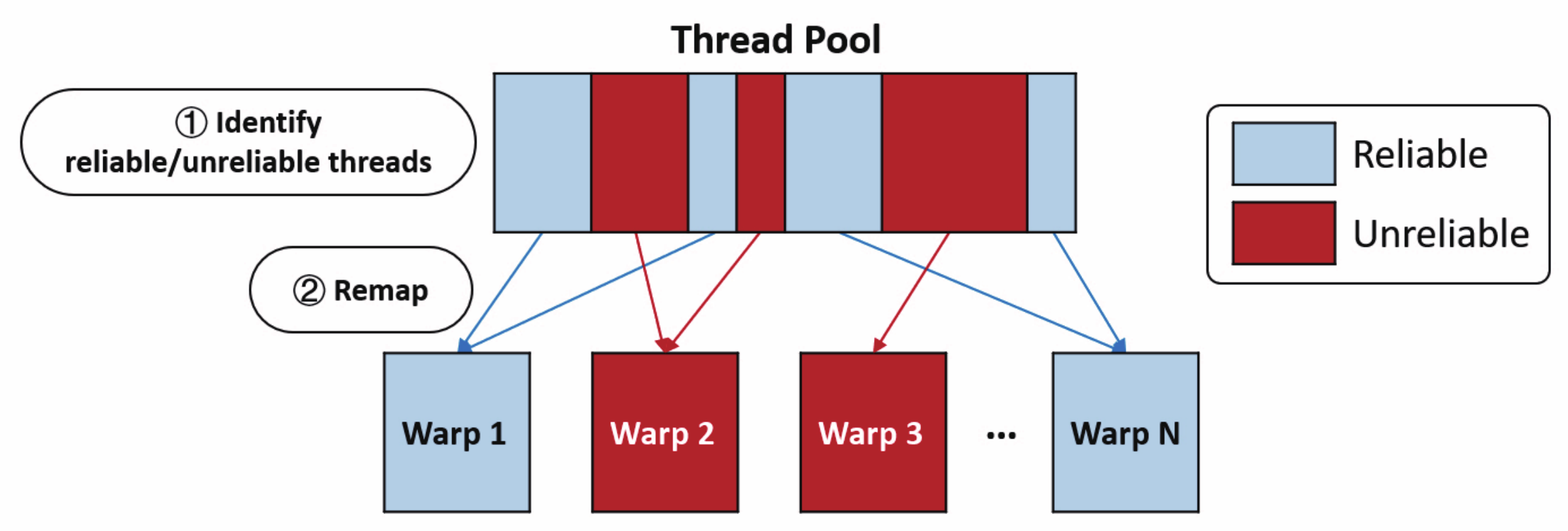}

		(b) Remapping.
	\end{minipage}

\vspace{\baselineskip}
	\begin{minipage}[t]{\columnwidth}
		\centering
		\includegraphics[scale=0.38]{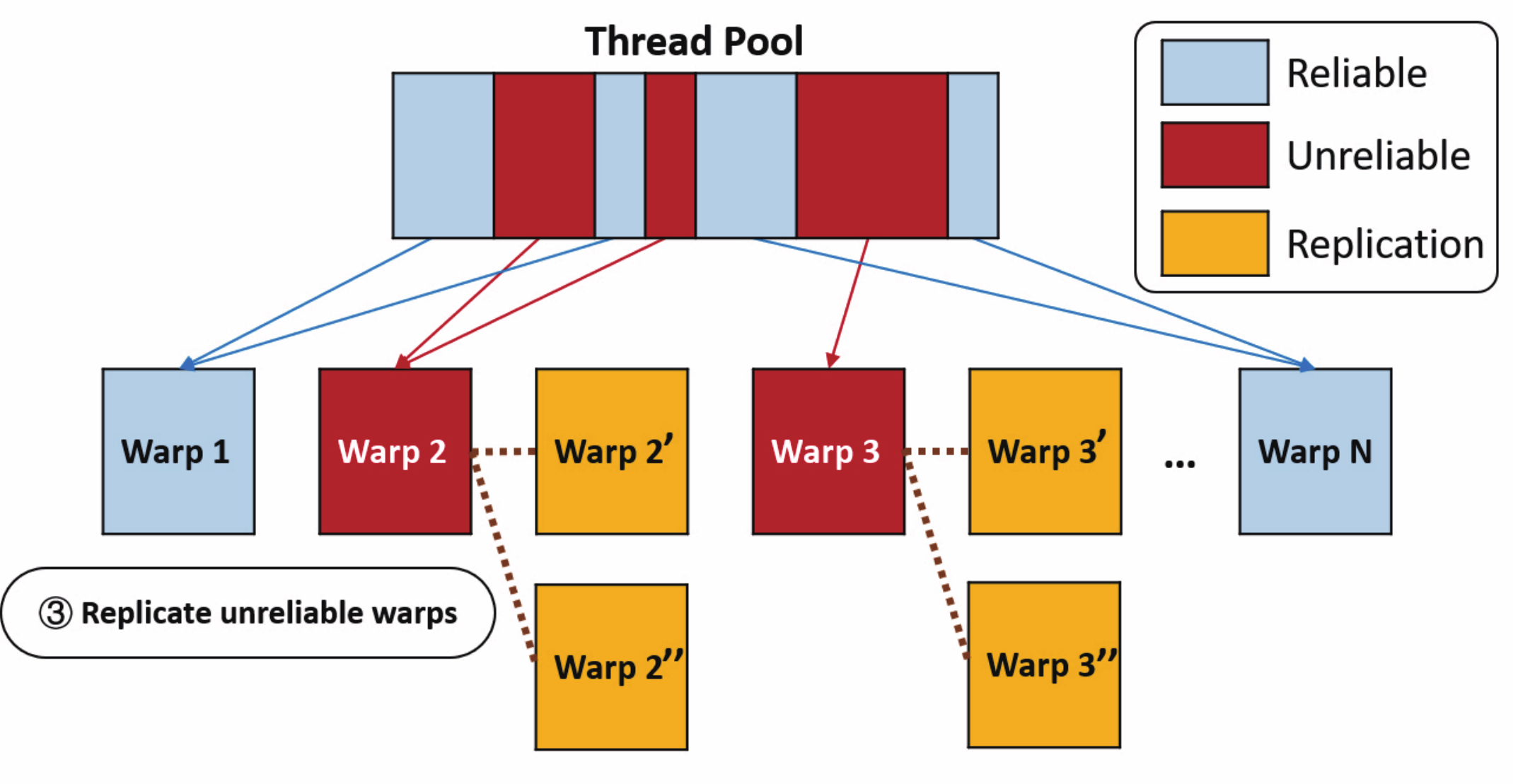}

		(c) Remapping and protection. 
		
		Unreliable warps are replicated once for detection. 
		
		If applying error correction, there are two replicas.
	\end{minipage}

	\caption{Logic of mapping, remapping, and protection.}
	\label{fig:remapping-logic}
\end{figure}

\subsection{Remapping}
\label{sec:remapping-logic}

CTAs are collections of threads defined by the CUDA programmer. The thread to warp mapping is done linearly by default (i.e., threads are allocated to warps in groups of 32 as shown in Figure~\ref{fig:remapping-logic}(a)). 
To change the mapping between thread to warp, we first identify reliable/unreliable threads in offline analysis, then remap threads, see Figure~\ref{fig:remapping-logic}(b).
By changing the linear thread order,
we can group threads with the same resilience (i.e., percentage of SDC outputs) into the same warp,
and use different resilience protection at the warp level according to their reliability profile. 
Note that  remapping is done within each CTA, but not across CTAs. This is because synchronization is ensured inside each CTA. Remapping across different CTAs can affect their synchronization, hence introduce errors in the software logic.

We use GPGPU-Sim~\cite{gpgpu-sim} to implement the above.
At the initialization phase, CTAs are constructed.
Without remapping, threads are launched linearly.
With remapping, we fill each CTA according to its resilience-based launching order.
This resilience-based launching order is decided based on offline profiling.
We start fetching threads from the beginning of the linear launching order, and put reliable threads into a warp.
Meanwhile, we organize unreliable threads into another warp, if any.
When a warp is filled (32 threads), the warp is ready for execution, and a new warp is formed for the upcoming threads.
When all the threads are remapped into warps, if there are any partially filled reliable or unreliable warps, they are combined into one mixed warp and ready for execution.
It is important to note that thread remapping to different warps does not affect their reliability profile because  thread resilience is typically determined by branch divergence and input data and not the order of thread execution~\cite{li2018modeling}.

\subsection{Partial Protection}
\label{sec:duplication}

In addition to remapping,  error detection/correction can be applied to unreliable warps.
Here, we use warp replication/triplication, to demonstrate how partial protection works.
During remapping, when an unreliable warp is filled (32 threads), we replicate it into another warp and send both warps to execute. 
After these two warps finish execution, thread outputs (usually the outputs are the computation results to be written into memory  by  {\sl store} instructions) are compared to detect whether there is any difference, see Figure~\ref{fig:remapping-logic}(c).
Since warps are the smallest unit for scheduling at the GPU level, duplication at the warp level 
is transparent for the programmer to handle than at the thread level.
Duplication at the CTA level would require redoing the logic of communication/synchronization among threads, a far more challenging software effort. This is fully avoided by handling replication at the warp level.

If error correction is applied, then each unreliable warp is triplicated, according to triple modular redundancy (TMR).
In Figure~\ref{fig:remapping-logic}(c), warp-2 is triplicated into warp-2' as well as warp-2" for error correction.
Since reliable warps do not need error detection/correction, they are not replicated/triplicated.
Note that, mixed warps that have both reliable and unreliable threads also need to be duplicated or triplicated for error detection/correction.

\begin{figure*}[t]
 	\centering
 	 	\begin{minipage}{\columnwidth}
 		\centering
 		\includegraphics[scale=0.4]{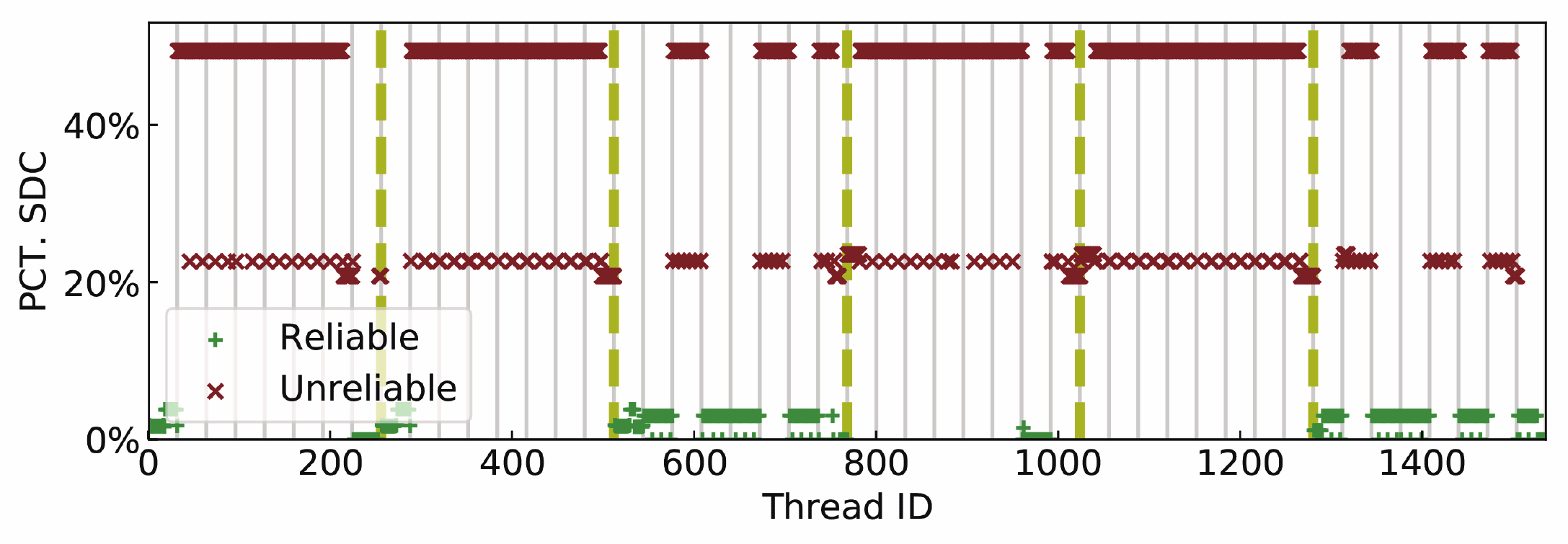}

         (a) HotSpot (first 6 CTAs).
 	\end{minipage}
  	\begin{minipage}{\columnwidth}
 		\centering
 		\includegraphics[scale=0.4]{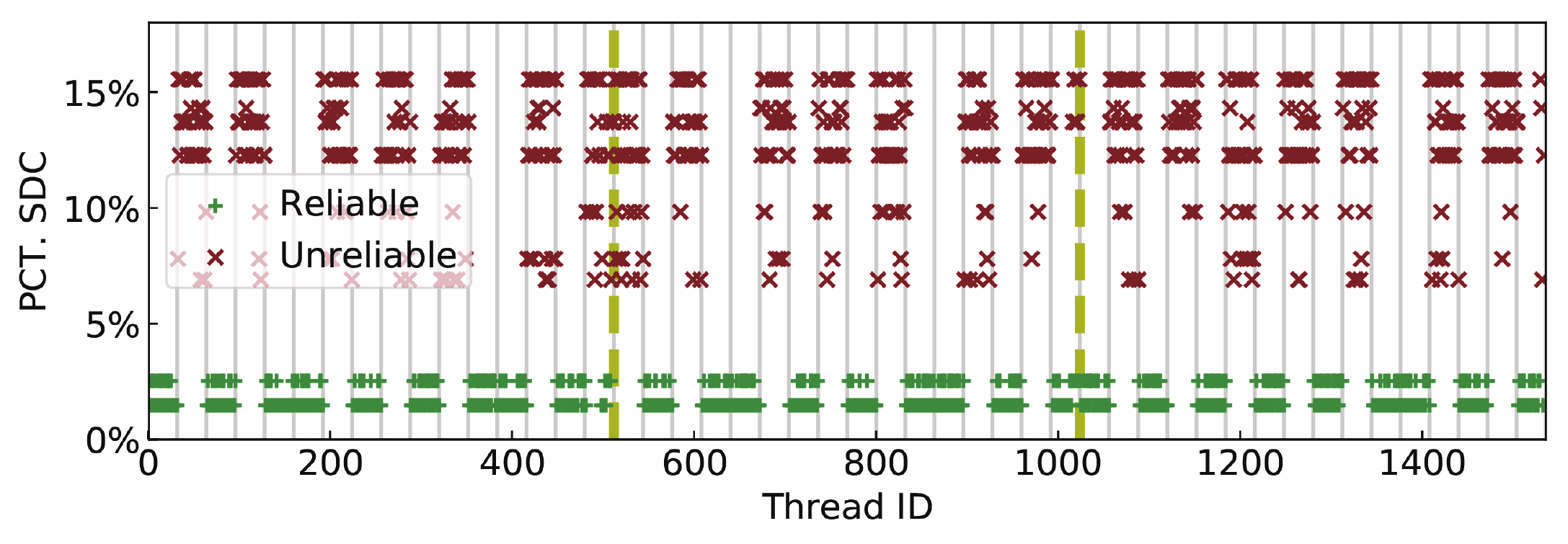}

         (b) Jmeint (first 3 CTAs).
 	\end{minipage}
 	\vspace{0.5mm}
 	
 	   	 	\begin{minipage}{\columnwidth}
 		\centering
 		\includegraphics[scale=0.4]{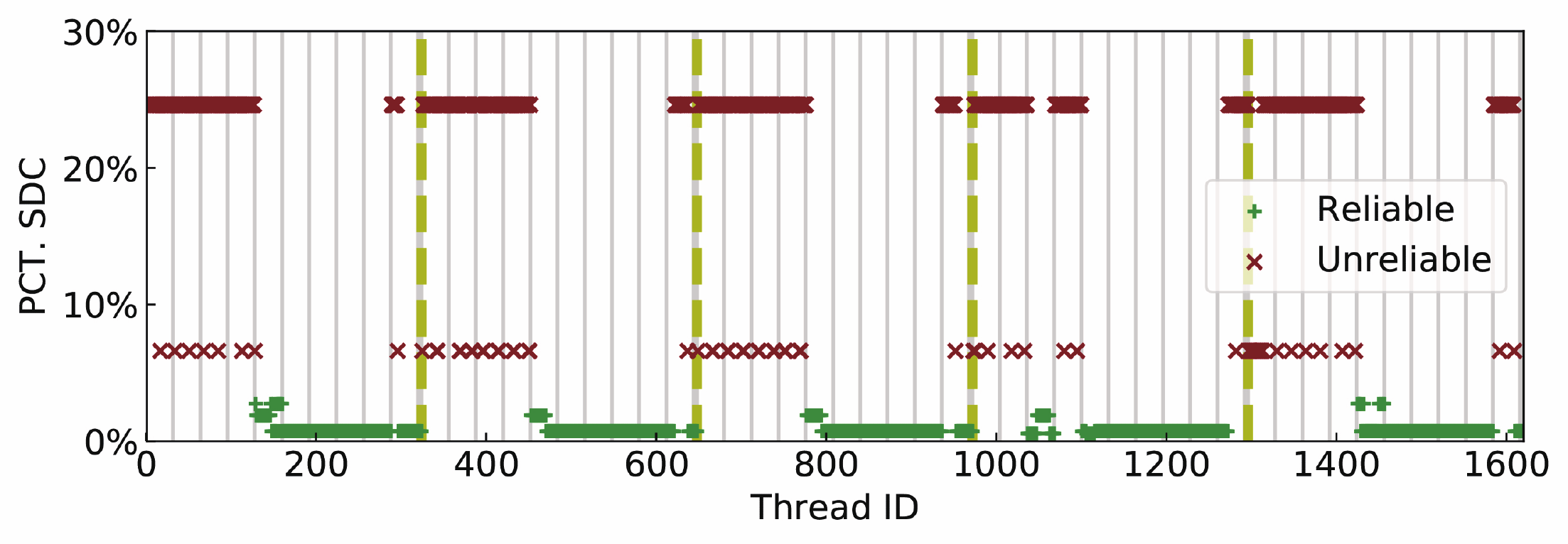}

         (c) Laplacian (first 5 CTAs).
 	\end{minipage}
 	 	 	\begin{minipage}{\columnwidth}
 		\centering
 		\includegraphics[scale=0.4]{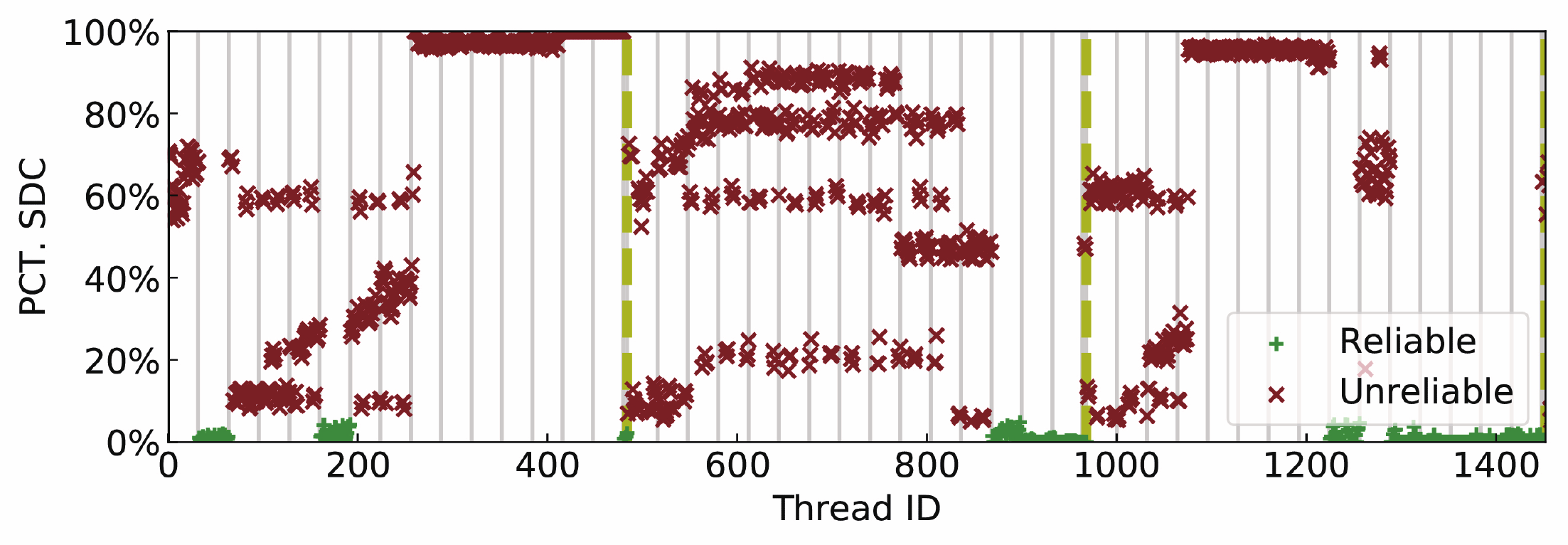}

         (d) MeanFilter (first 3 CTAs).
 	\end{minipage}
 	
   	\vspace{0.5mm}
 	 	\begin{minipage}{\columnwidth}
 		\centering
 		\includegraphics[scale=0.4]{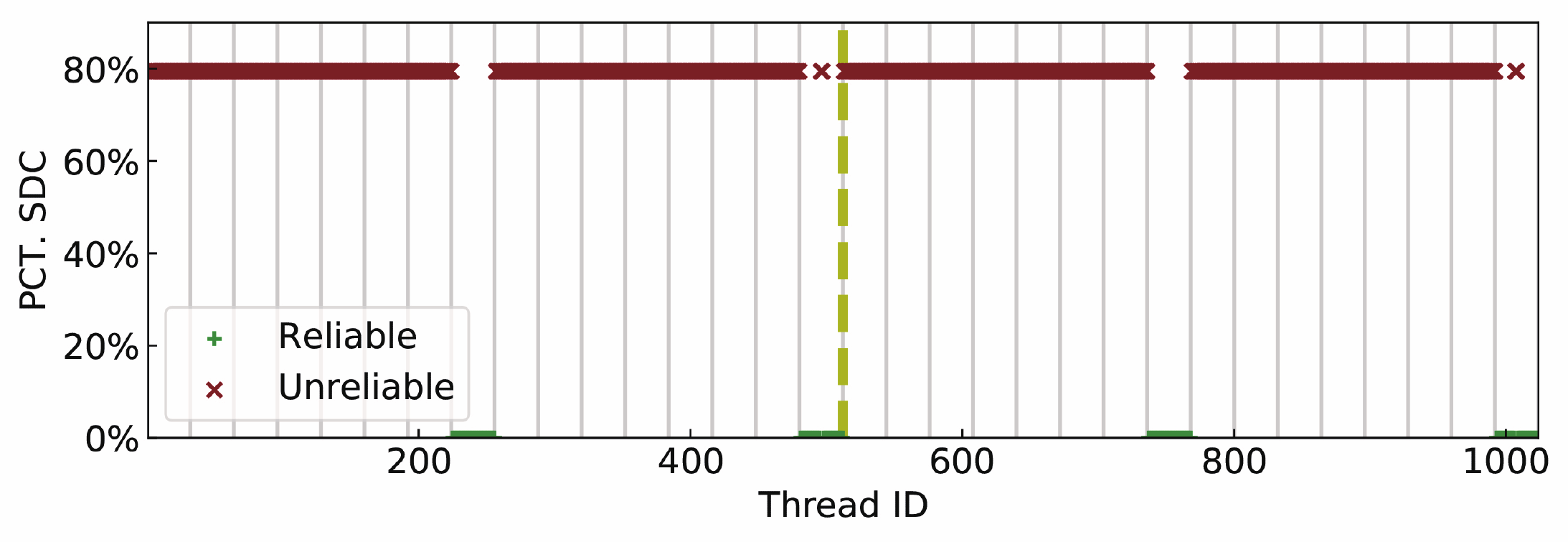}
         (e) 2DCONV.
 	\end{minipage}
  	 	 	\begin{minipage}{\columnwidth}
 		\centering
 		\includegraphics[scale=0.4]{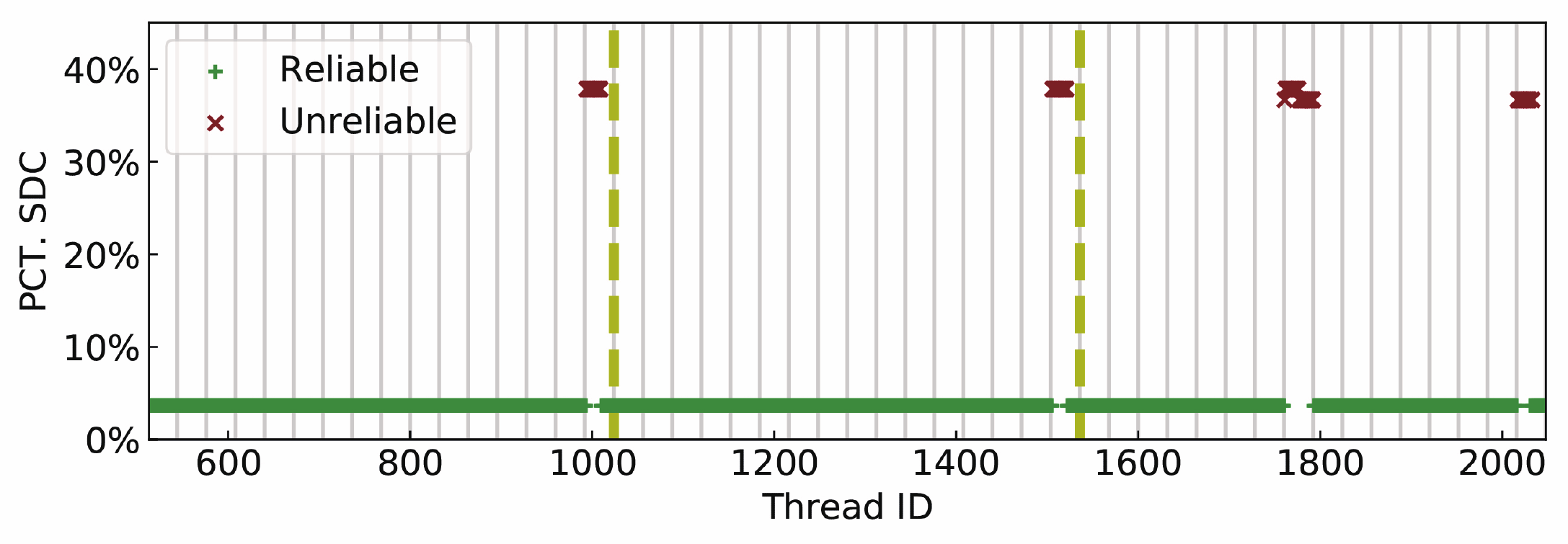}

         (f) Gaussian~K2 (first 3 CTAs).
 	\end{minipage}	
\vspace{0.5mm}

 	 	\begin{minipage}{\columnwidth}
 		\centering
 		\includegraphics[scale=0.4]{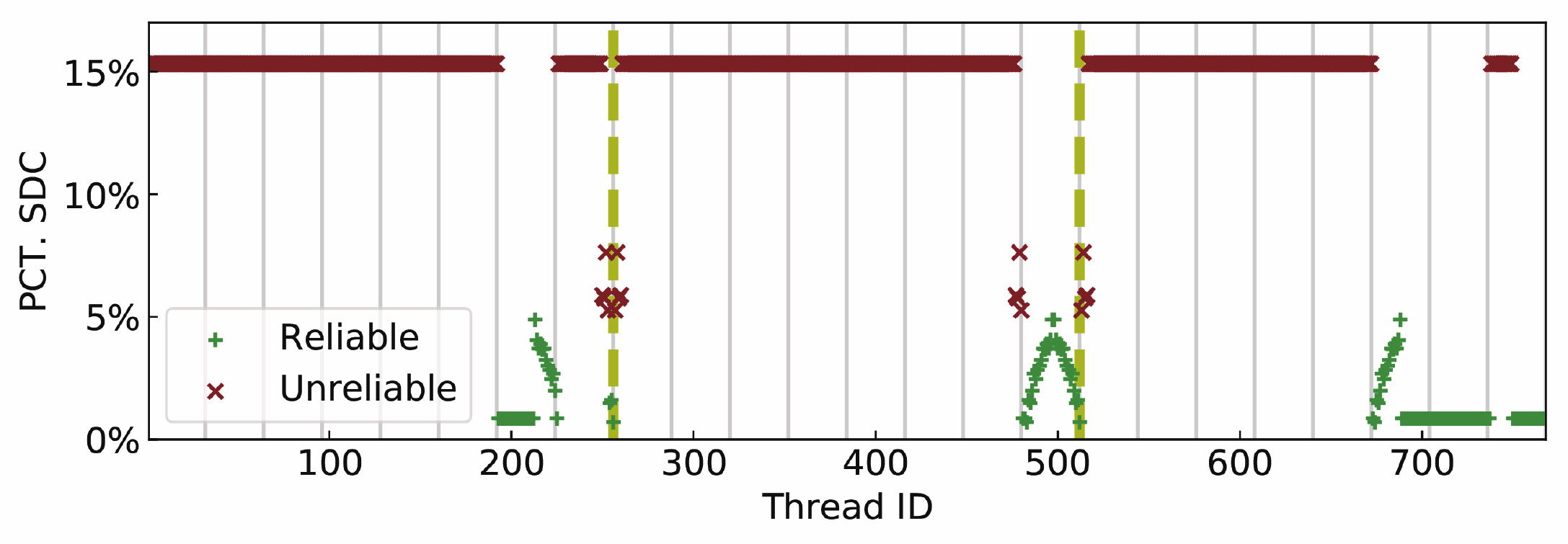}

         (g) PathFinder.
 	\end{minipage}

 	\caption{Resilience patterns after remapping. If a thread has SDC probability less or equal to 5\%, it is considered reliable. Due to space constraint, we only show the first several CTAs for HotSpot, Jmeint, Laplacian, MeanFilter, and Gaussian~K2.}
 	\label{fig:benchmark-remapping}

 \end{figure*}

 \begin{figure}[htb]
 	\centering
 	\begin{minipage}{\columnwidth}
 		\centering
 		\includegraphics[scale=0.4]{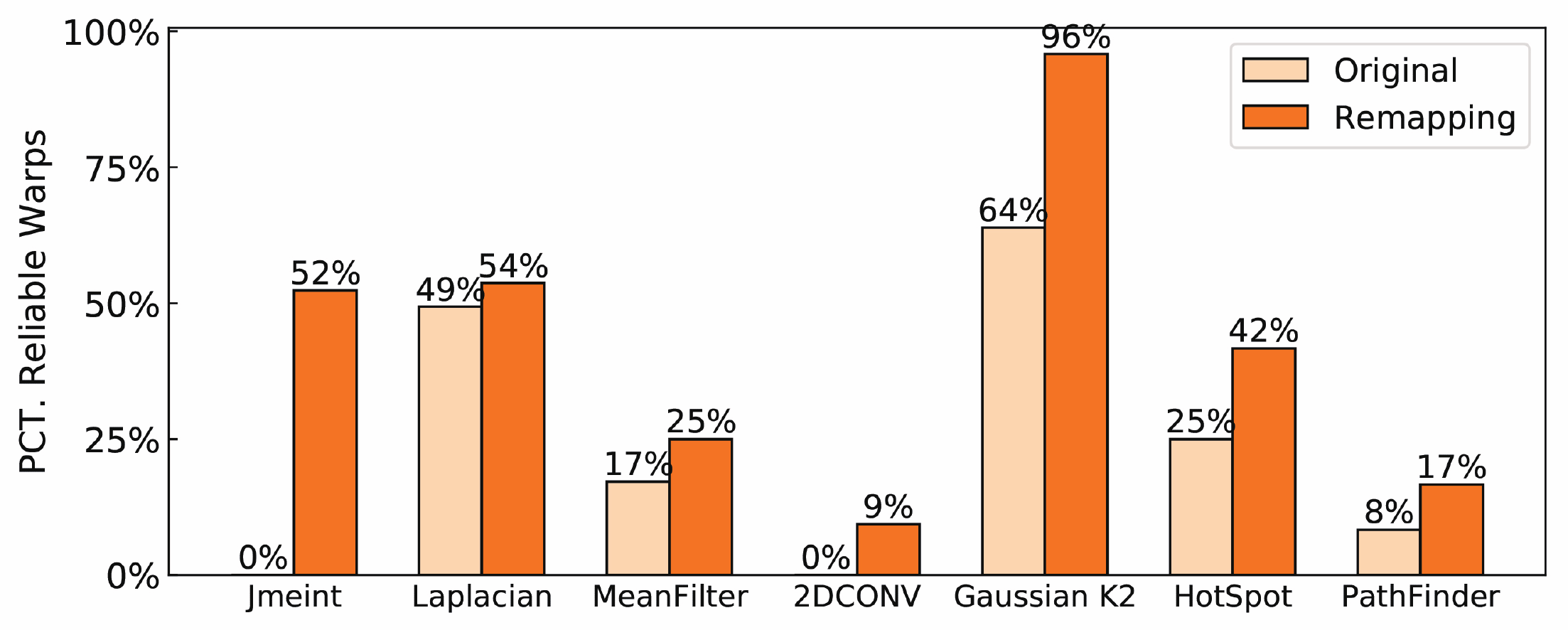}
 	\end{minipage}
  	\caption{Percentage of reliable warps before and after remapping. }
 	\label{fig:pct_warp}
 \end{figure}

\section{Evaluation}
\label{sec:evaluation}

In this section we present a detailed evaluation of thread remapping (Section~\ref{sec:remapping-effectiveness}). Then, we discuss the  overhead magnitude when applying protection via remapping (Section~\ref{sec:remapping-overhead}).

\subsection{Effectiveness of Thread Remapping}
\label{sec:remapping-effectiveness}

We first show the resilience pattern of different benchmarks after remapping, see Figure~\ref{fig:benchmark-remapping}.
 In this figure, if a thread has an SDC probability less or equal to 5\%, it is considered reliable, and remapping is performed based on this 5\% SDC threshold, i.e., our goal is a 95\% reliability coverage.
Gray solid lines in Figure~\ref{fig:benchmark-remapping} separate different warps, and yellow dashed lines separate different CTAs.
 For HotSpot, in the first CTA, originally all the reliable threads in Figure~\ref{fig:reliable_unreliable_together}(c) are distributed across all warps.
 After remapping, these reliable threads are gathered and scheduled in the first and last warp of the CTA.
 We end up with 1, 5, 1, and 5 reliable warps for the second, third,  fourth, and sixth CTA, respectively.
 There is a mixed warp at the end of the third CTA. Since there are still unreliable threads in this mixed warp,  it still needs protection.
 For the fifth CTA, all threads are unreliable, therefore 
the thread resilience pattern is the same before and after remapping.
The resilience patterns after remapping for Jmeint, Laplacian, MeanFilter, 2DCONV, Gaussian~K2, and PathFinder are shown in
Figure~\ref{fig:benchmark-remapping}(b)-(f).

The improvement in terms of the percentage of reliable warps for applications is shown in Figure~\ref{fig:pct_warp}.
On average, originally the percentage of reliable warps is 23.40\%. By remapping, the percentage increases to 42.08\%.
The biggest improvement happens on Jmeint, where there are 52\% reliable warps after remapping from the original 0\%.

 \begin{figure}[tb]
 	\centering
  	\begin{minipage}{\columnwidth}
 		\centering
 		\includegraphics[scale=0.4]{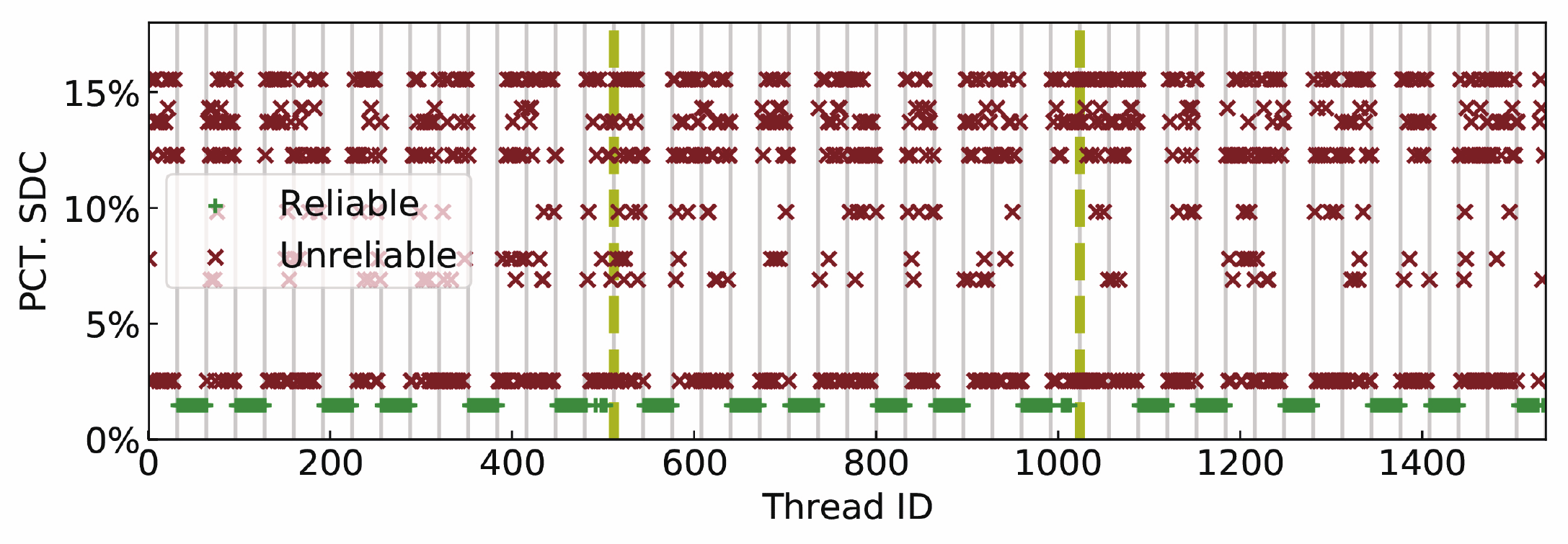}

         (a) threshold = 2\%
 	\end{minipage}
 	
 	\vspace{1mm}
 	 	\begin{minipage}{\columnwidth}
 		\centering
 		\includegraphics[scale=0.4]{scatter_resilience_remapping_jmeint_d4096a_k1_threshold_05.pdf}

         (b) threshold = 5\%
 	\end{minipage}
 	\vspace{1mm}
 	 	\begin{minipage}{\columnwidth}
 		\centering
 		\includegraphics[scale=0.4]{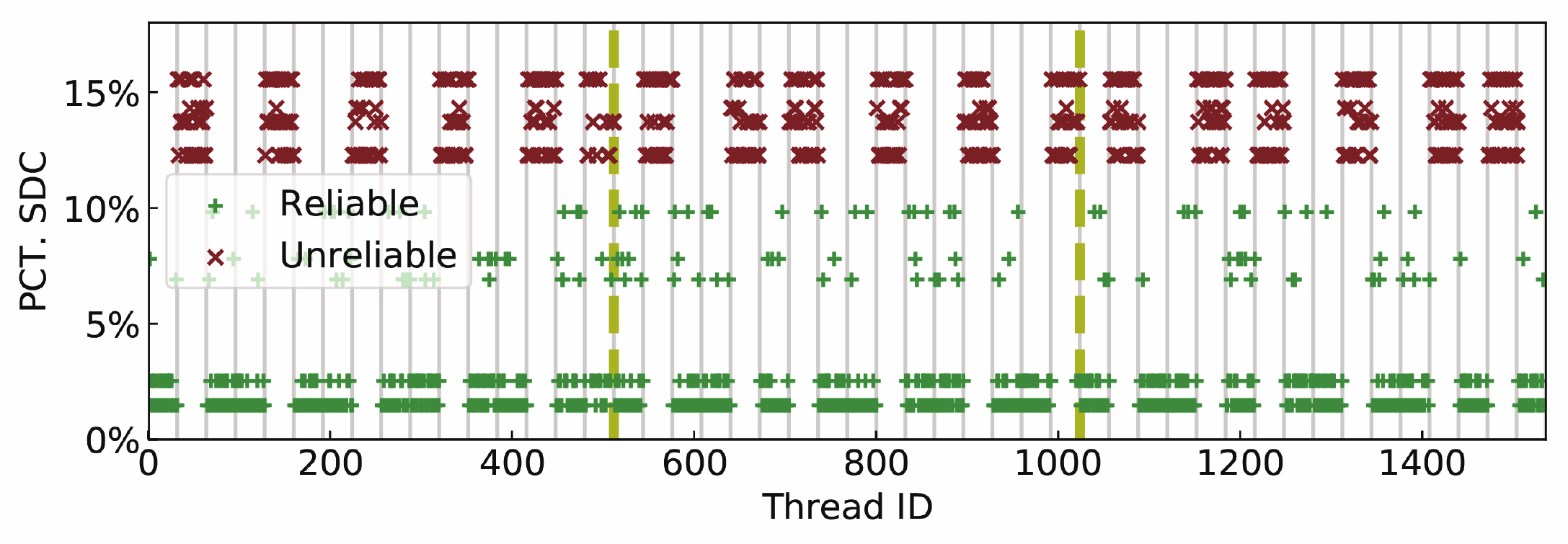}

         (c) threshold = 12\%
 	\end{minipage}
 	\vspace{1mm}
 	 	\begin{minipage}{\columnwidth}
 		\centering
 		\includegraphics[scale=0.4]{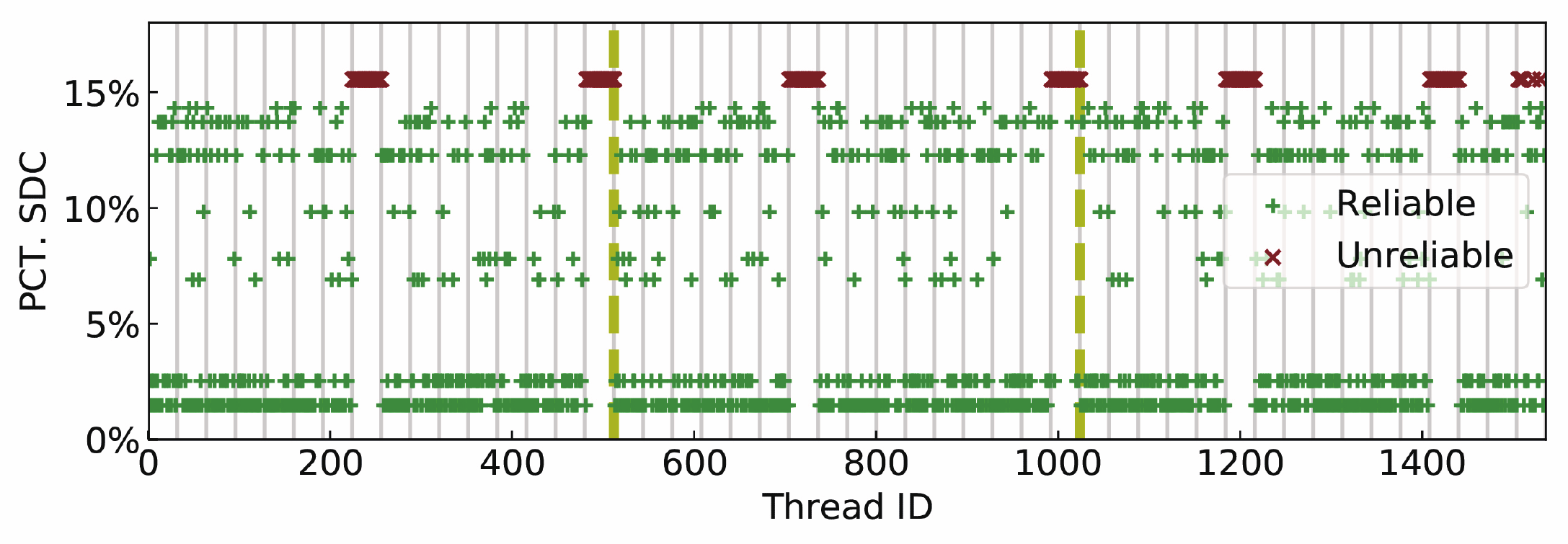}

         (d) threshold = 15\%
 	\end{minipage}
 	\vspace{1mm}
 	\caption{Remapped resilience patterns of Jmeint under different SDC threshold. Due to space constraint, we only show the first 3 CTAs.}
 	\label{fig:jmeint-remapping-threshold}

 \end{figure}

 \begin{figure}[htb]
 	\centering
 	\begin{minipage}{\columnwidth}
 		\centering
 		\includegraphics[scale=0.4]{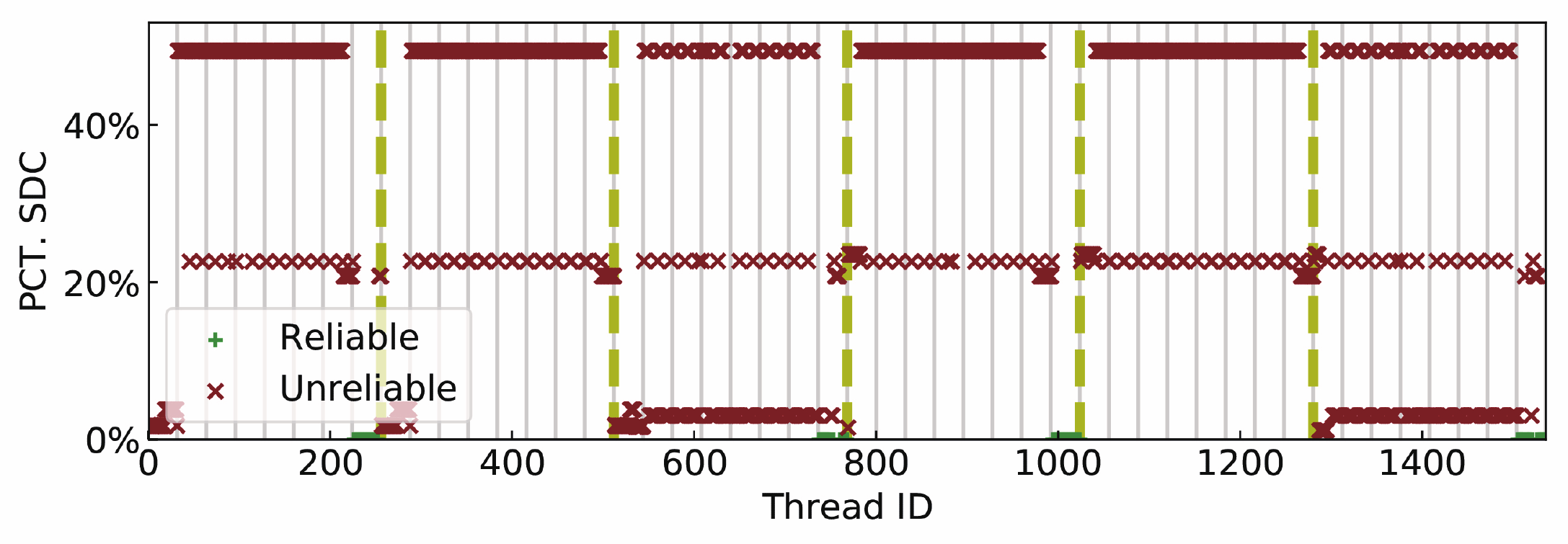}

         (a) threshold = 0\%
 	\end{minipage}
 	\vspace{1mm}
 	 	\begin{minipage}{\columnwidth}
 		\centering
 		\includegraphics[scale=0.4]{scatter_resilience_remapping_hotspot_d32_k1_threshold_05.pdf}

         (b) threshold = 5\%
 	\end{minipage}
 	\vspace{1mm}
 	 	\begin{minipage}{\columnwidth}
 		\centering
 		\includegraphics[scale=0.4]{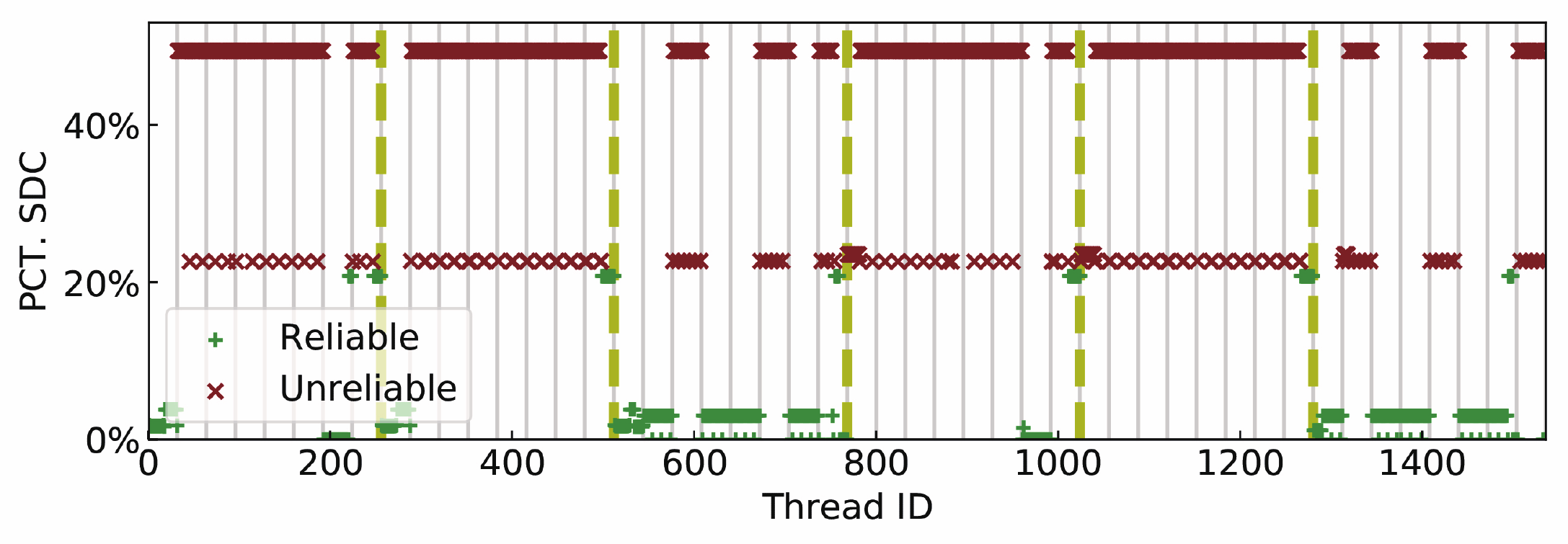}

         (c) threshold = 22\%
 	\end{minipage}
 	\vspace{1mm}
 	 	\begin{minipage}{\columnwidth}
 		\centering
 		\includegraphics[scale=0.4]{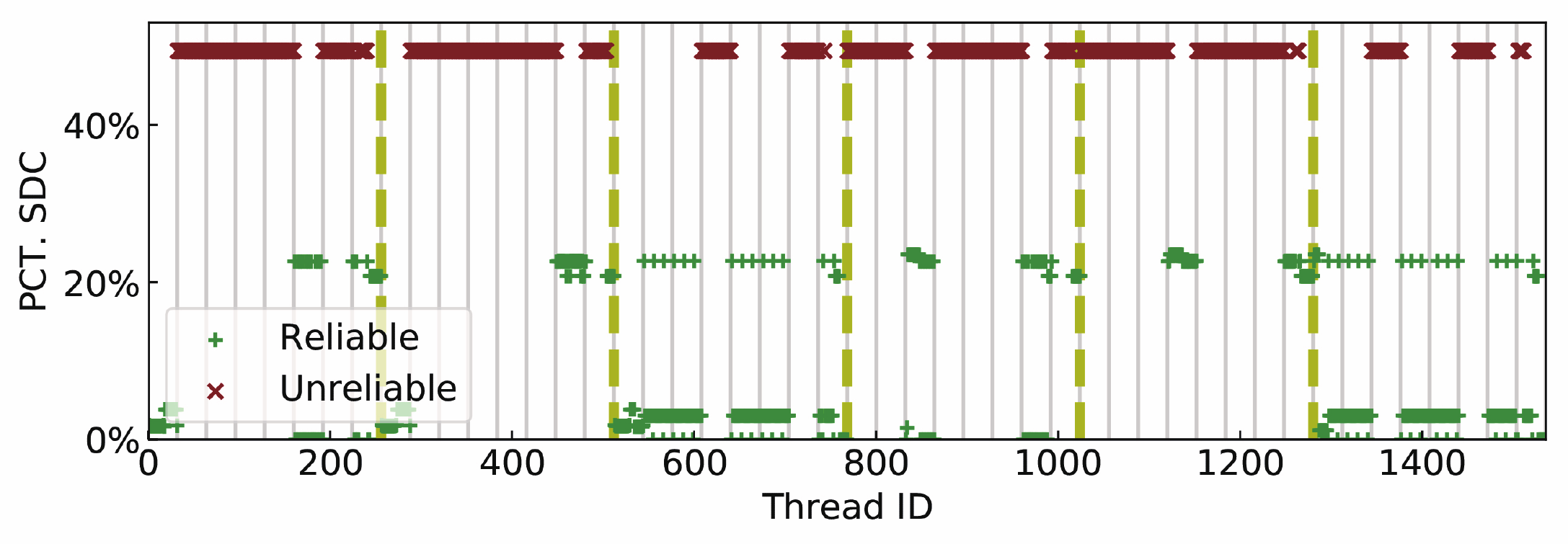}

         (d) threshold = 25\%
 	\end{minipage}
 	\vspace{1mm}
 	\caption{Remapped resilience patterns of HotSpot under different SDC threshold.  Due to space constraint, we only show the first 6 CTAs.}
 	\label{fig:hotspot-remapping-threshold}

 \end{figure}

   \begin{figure}[!htb]
  	\centering
  	\begin{minipage}{1\columnwidth}
  		\includegraphics[scale=0.4]{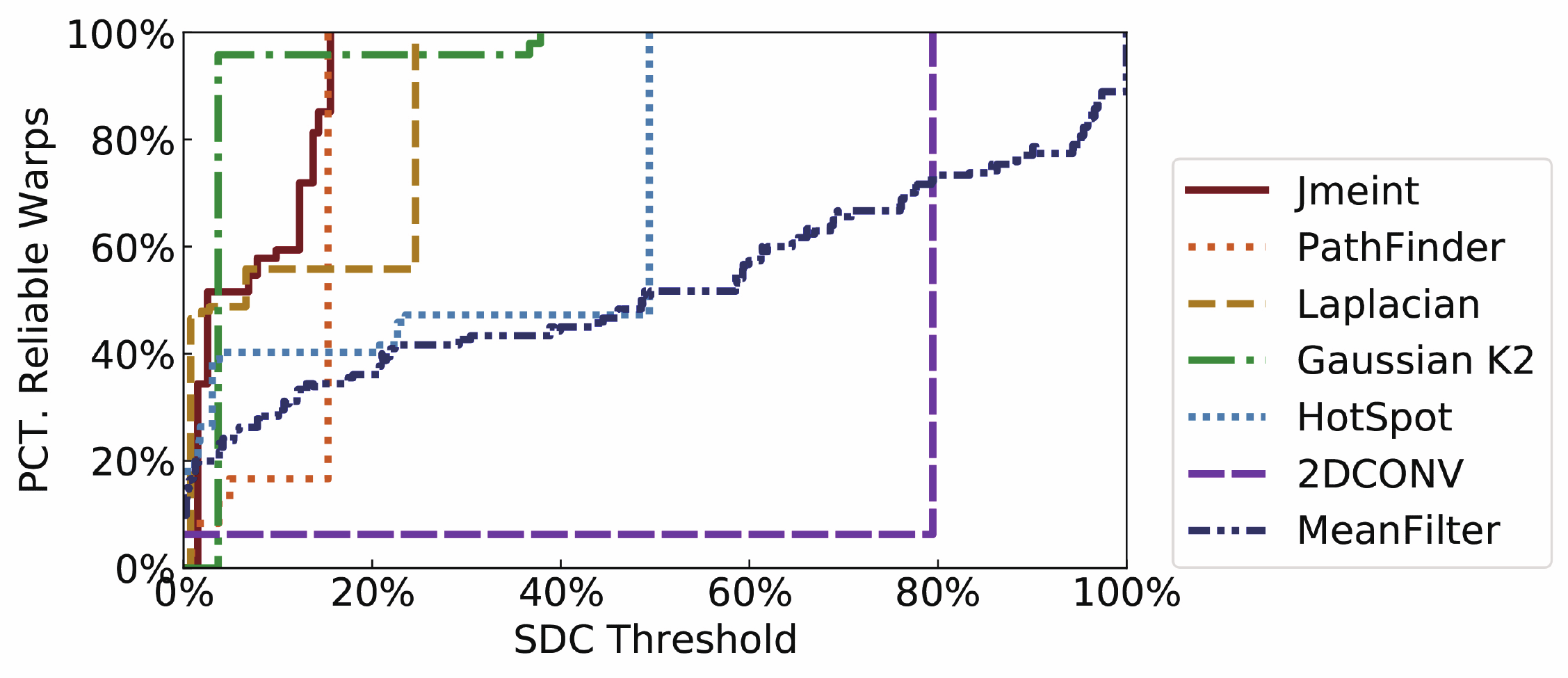}
  	\end{minipage}
  	\caption{Percentage of reliable warps grows as the SDC tolerance threshold increases.}
  	\label{fig:sdc-threshold-to-pct-warp}
  \end{figure}

Changing the SDC tolerance threshold can result in different remappings.
Figure~\ref{fig:jmeint-remapping-threshold} shows how remapping changes the resilience pattern when different SDC thresholds are applied in Jmeint.
If the SDC threshold is set to 2\%, there are a few reliable threads to be remapped, see Figure~\ref{fig:jmeint-remapping-threshold}(a).
For increased SDC thresholds, remapping results in more reliable warps, see the changes of resilience patterns in Figure~\ref{fig:jmeint-remapping-threshold}(a) to (d).

For Hotspot, even for SDC threshold equal to 0\%, there are still several reliable warps (two warps in  Figure~\ref{fig:hotspot-remapping-threshold}(a) within the first 6 CTAs and in total 25\% for the whole kernel).
With the SDC threshold increasing, see Figure~\ref{fig:hotspot-remapping-threshold}(b)-(d), remapping changes the resilience pattern, and
more reliable threads are gathered together.

Figure~\ref{fig:sdc-threshold-to-pct-warp} shows how the percentage of SDC outputs changes when the SDC threshold increases for all 7 benchmarks eligible for remapping.
Jmeint and PathFinder are the first two benchmarks reaching 100\% reliable warps, with SDC threshold less than $20\%$.
Gaussian~K2 has 94\% reliable warps when the SDC threshold is 3.6\% only. This is because the SDC percentage of its major thread group is 3.6\%.
HotSpot reaches 100\% reliable when SDC threshold is about 50\%, and 2DCONV requires SDC threshold to be 80\% to get 100\% reliable warps.
MeanFilter is the most complicated benchmark, and it reaches 100\% reliable only when SDC threshold is set to 100\%, because 15\% of the threads have 100\% SDC outputs.
In general, we see that if we set the SDC threshold to 5\% only, there is still ample room for remapping for most benchmarks, as shown in Figure~\ref{fig:pct_warp}.

 \begin{figure}[tb]
 	\centering
 	\begin{minipage}{\columnwidth}
 		\centering
 		\includegraphics[scale=0.4]{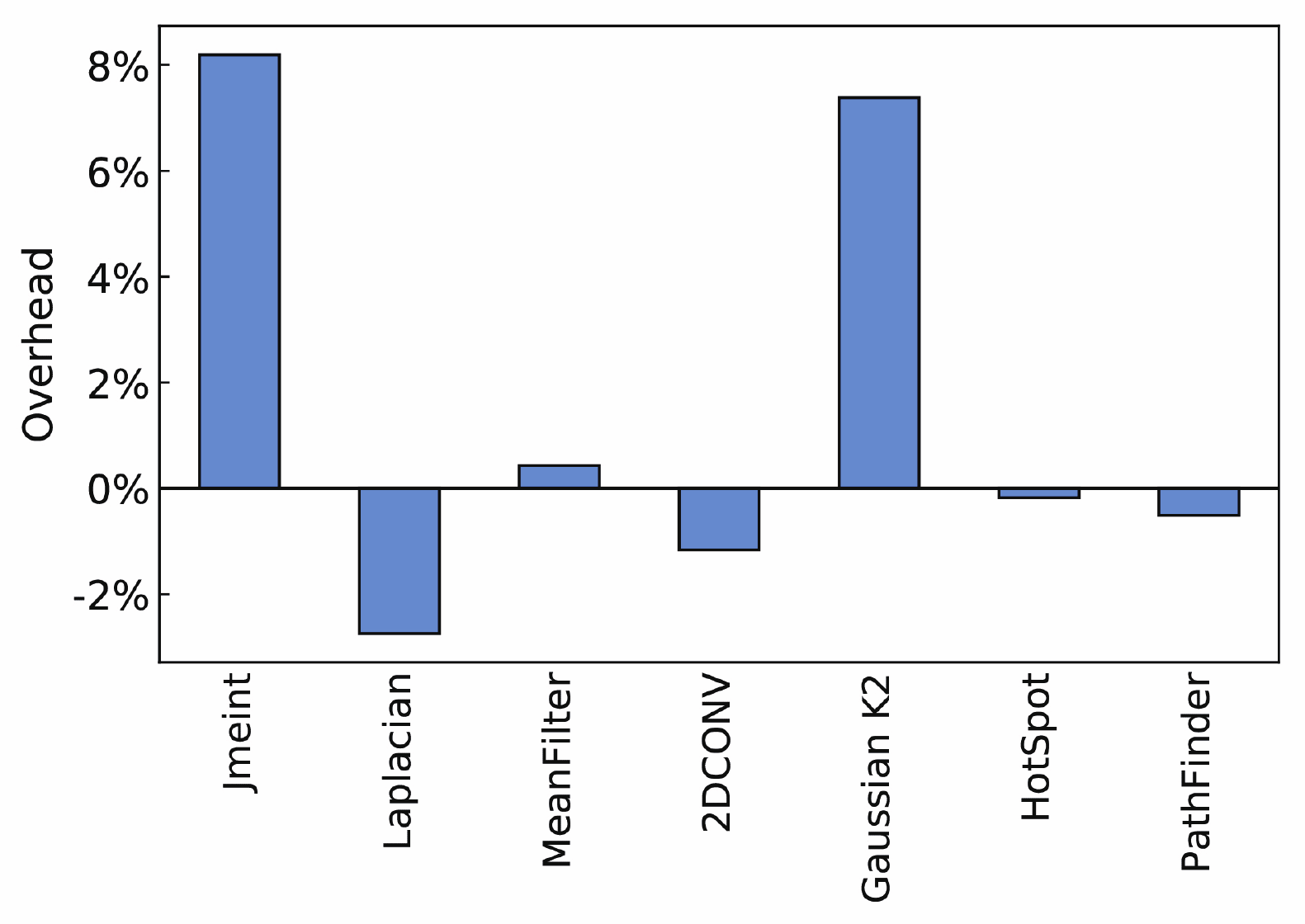}
 	\end{minipage}
  	\caption{Overhead of remapping. }
 	\label{fig:rm_overhead}
 \end{figure}

 \begin{figure}[tb]
 	\centering
 	\begin{minipage}{0.48\columnwidth}
 		\centering
 		\includegraphics[scale=0.3]{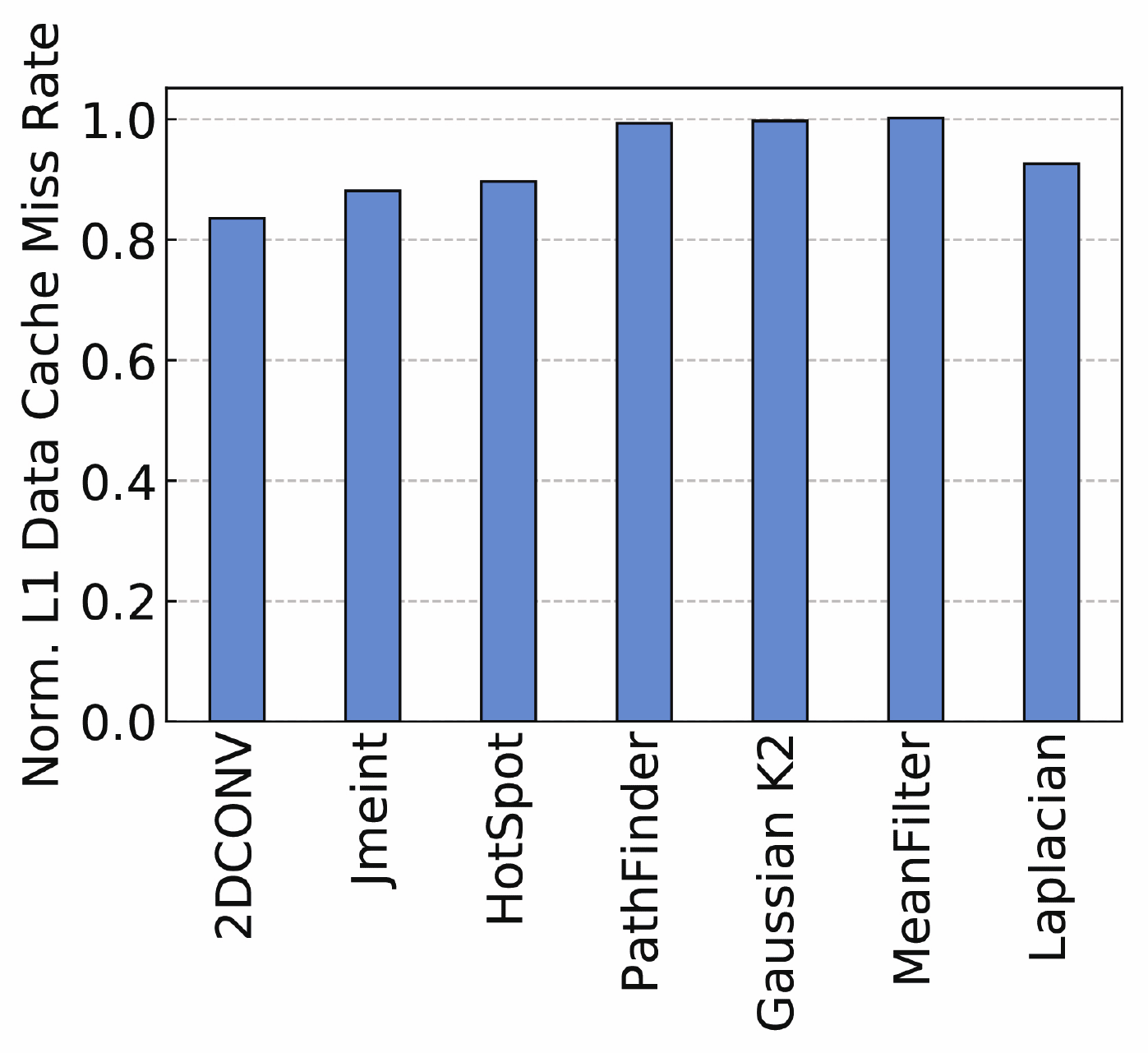}

 		(a)  L1 Data Cache Miss Rate
 	\end{minipage}
 	\begin{minipage}{0.48\columnwidth}
 		\centering
 		\vspace{0.3cm}
 		\includegraphics[scale=0.3]{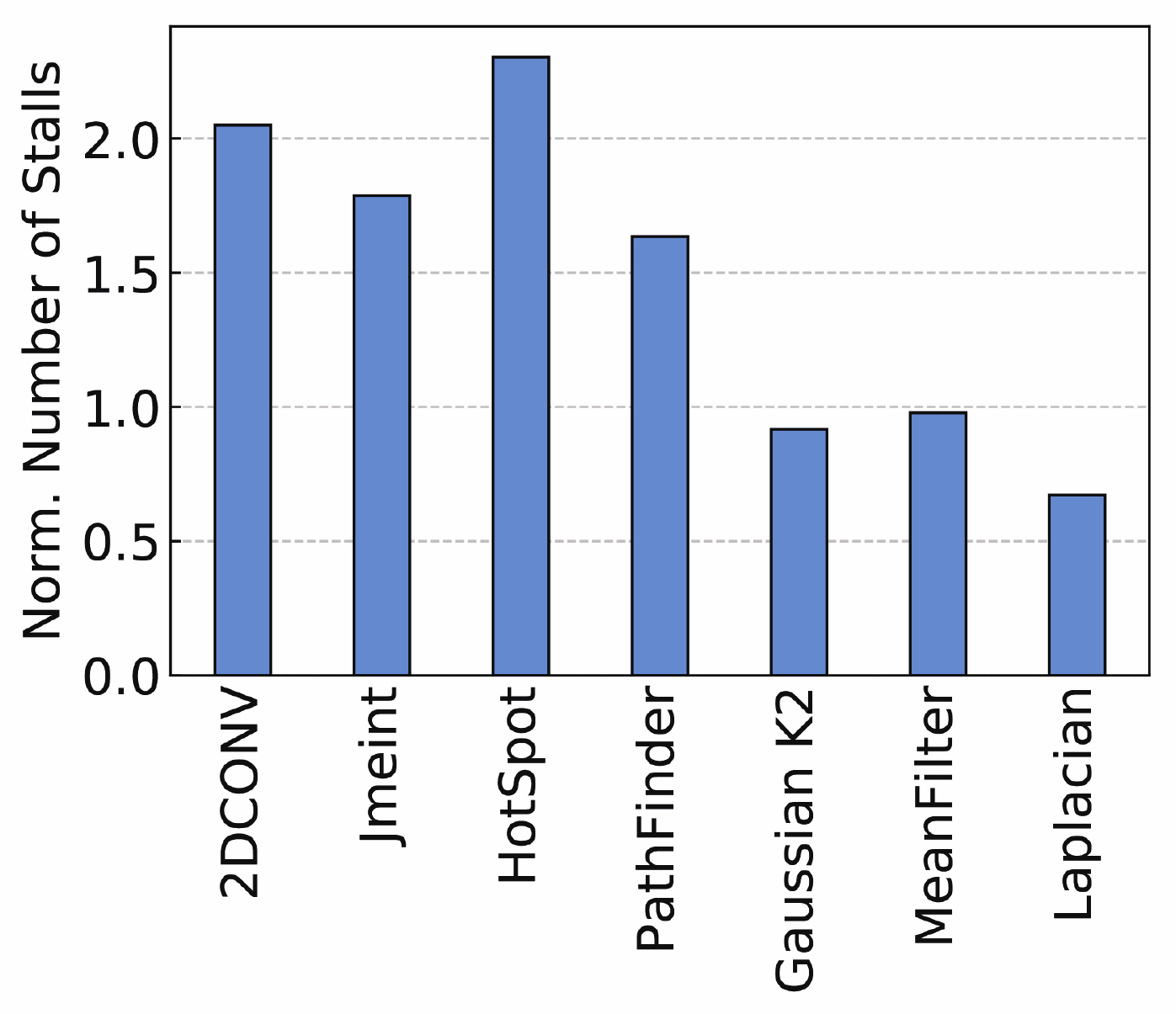}

 		(b) Number of Stalls
 	\end{minipage}
  
  	\caption{Detailed metrics of remapping overhead. All numbers are normalized  by the execution without remapping. }
 	\label{fig:detail_res}
 \end{figure}

\subsection{Overhead Introduced by Remapping and Protection}
\label{sec:remapping-overhead}

Thread remapping (and protection) may affect the performance of program execution.
Because of the  shared cluster environment we are using, pure timing measurement is not accurate. Instead,
 we use the number of instruction cycles measured using GPGPU-Sim performance mode
to reflect the execution performance.

For overhead analysis, we first present the  performance overhead due to remapping. 
The remapping overhead of each benchmark kernel 
is shown in Figure~\ref{fig:rm_overhead}. 
On average, the remapping overhead is only 1.63\%.
Note that there are some benchmarks with  negative overhead in Figure~\ref{fig:rm_overhead}, such as Laplacian, 2DCONV, HotSpot, and PathFinder, in these cases execution cycles reduce with remapping.

To better understand why remapping may result in better performance, we look into various performance measures that are normalized over the original thread mapping.
Figure~\ref{fig:detail_res} shows the normalized L1 data cache miss rate and the number of stalls caused by accessing shared memory.
The numbers are normalized by the execution without remapping.
On the one hand, from Figure~\ref{fig:detail_res}(a), we observe that the L1 data cache miss rate is decreasing. On the other hand, Figure~\ref{fig:detail_res}(b) shows that the number of stalls  increases for 2DCONV, Jmeint, HotSpot, and PathFinder; for Laplacian, MeanFilter, and Gaussian~K2, the number of stalls decreases.
Trends are not consistent across benchmarks, therefore some gain and some lose performance with remapping.
In sum, we claim that remapping does not significantly affect  performance which remains in the same ballpark as the original cases.

\begin{figure}[htb]
 	\centering
 	\begin{minipage}{\columnwidth}
 		\centering
 		\includegraphics[scale=0.4]{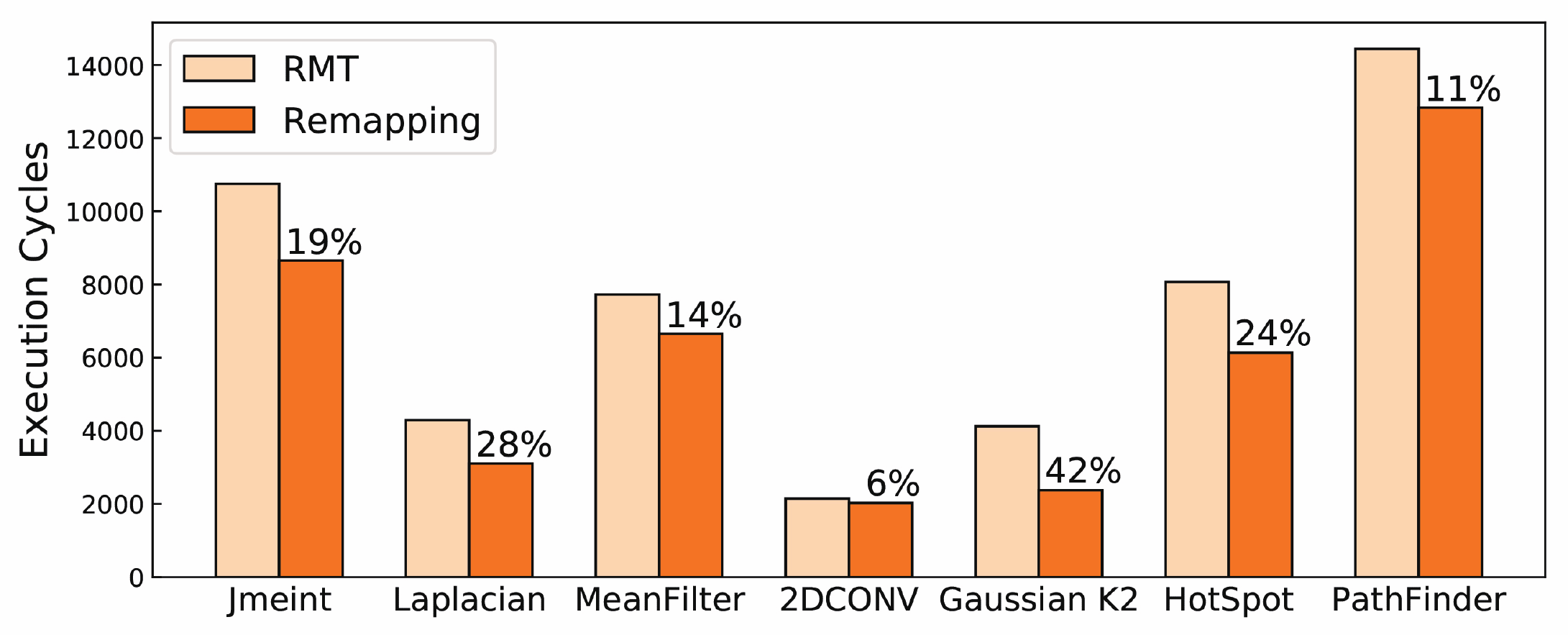}

 	\end{minipage}

  	\caption{Comparison of execution performance using duplication for error detection between remapping and RMT. 
  	}
 	\label{fig:speedup-rmt}
 \end{figure}

Last but not least, we show the performance savings of applying error detection/correction after remapping.
In the case of error detection, we compare our technique with RMT (Redundant Multi-Threading), where all the threads (all warps) are duplicated for error detection.
Figure~\ref{fig:speedup-rmt} shows the execution performance of our remapping technique in execution cycles, comparing to RMT.
We also present the percentage of saved execution cycles at the top of each application bar.
On average, the percentage of saved execution cycles for error detection is 20.61\%, while Gaussian~K2 achieves a significant 42.39\% savings.

\begin{figure}[htb]
 	\centering
 	\begin{minipage}{\columnwidth}
 		\centering
 		\includegraphics[scale=0.4]{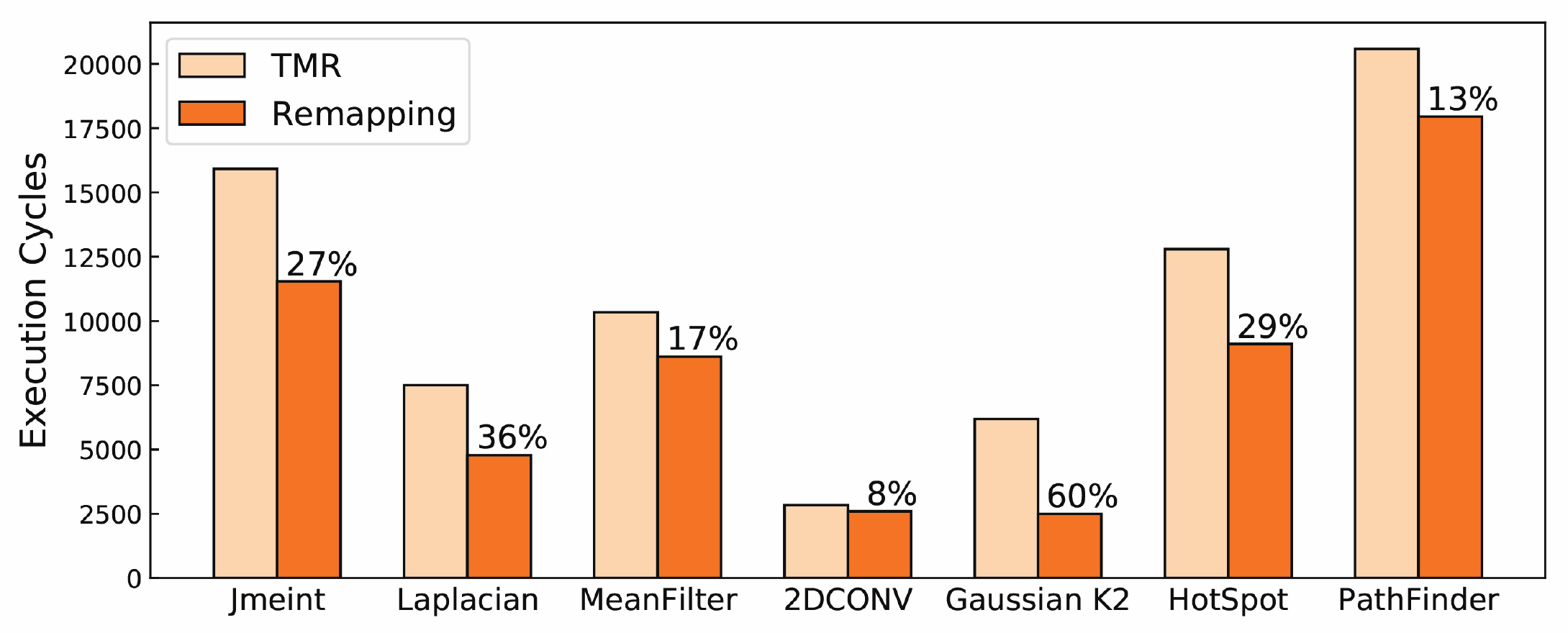}
 	\end{minipage}

 	\caption{Comparison of execution performance using triplication for error correction between remapping and TMR. }
 	\label{fig:speedup-tmr}
  	\vspace{0.4cm}
 \end{figure}

In addition, we compare  partial protection via remapping  with TMR (Triple Modular Redundancy), results are shown in Figure~\ref{fig:speedup-tmr}. 
The average saving in terms of execution cycles is 27.15\%, and again, Gaussian~K2 has the highest savings of 60.02\%.
Generally performance results are similar for both error detection and correction, and the saving of error correction  is always higher for every benchmark.
This is expected, since partial protection using triplication avoids the  execution  of two copies for all reliable warps, while for error detection with duplication, we only save one copy execution of reliable warps.
The per benchmark savings are  related to the benchmark resilience profile, i.e., the percentage of reliable threads in each CTA.
Since 95.87\% of threads in Gaussian~K2 is reliable, this benchmark achieves the highest savings.

{\sl Summary.} We show the effectiveness of remapping by analyzing the percentage of reliable warps, which increases on average from  23.40\% to 42.08\% with remapping.
Remapping introduces moderate to insignificant overhead. After applying remapping with protection,  an average saving of 20.61\% and 27.15\% execution cycles for error detection (duplication of CTAs without remapping) and correction (triplication of CTAs without remapping), respectively.

\section{Related Work}
\label{sec:related-work}

Several works address reliability within the software engineering domain. Chiminey~\cite{yusuf2015chiminey} provides a reliable platform for cloud computing.  
Bleser et al.~\cite{de2020delta} presents an automated approach to analyze the resilience of actor programs in distributed systems.
Chan et al.~\cite{chan2017ipa} uses  invariants to study error propagation in multi-threading applications using software fault injection.
Yang et al.~\cite{yang2020far} uses a software fault injection tool to evaluate different  anomaly detectors.
However, none of these works are applied in the context of GPUs.

Redundancy-based solutions are used to protect GPGPU applications from errors. Such solutions rely on double execution~\cite{dimitrov2009understanding,wadden2014real,mahmoud2018optimizing} for error detection, called \textit{dual-modular redundancy} (DMR) and triple execution~\cite{milluzzi2017exploration,chen2016gpu} for error correction, called \textit{triple-modular redundancy} (TMR).
Dimitrov et al.~\cite{dimitrov2009understanding} first evaluate the overhead of introduced redundancy at kernel level, thread level, and instruction level and show that at all levels the overhead can be over 90\%.
 Wadden et al.~\cite{wadden2014real} take a deeper look at two different ways of applying redundant multithreading (RMT) at the granularity of CTAs (i.e., intergroup RMT and intragroup RMT) and present the trade-off between overhead and resilience coverage.
Mahmoud et al.~\cite{mahmoud2018optimizing} choose instruction-level redundancy as it is transparent to programmers and propose SInRG, a collection of several software and hardware optimizations, to further reduce overhead.
In addition to those works targeting error detection, researchers also propose various solutions to correct errors with reduced overhead~\cite{milluzzi2017exploration,chen2016gpu} as compared to a naive implementation of TMR with triple overhead.

While the aforementioned solutions all focus on comparing and analyzing various redundancy-based protection solutions and seeking opportunities to reduce redundancy overhead, the ``partial protection" methodology approaches this problem from a totally different perspective by focusing on reducing the portion of threads that require protection and on organizing the threads in such a manner that result in more reliable software.

\section{Conclusions}
\label{sec:conclusion}

We presented a methodology to remap threads into warps according to their resilience profile. Looking into 12 benchmarks (17 kernels) from four benchmark suites, we identified that 7 of them are amenable to remapping for resilience. The proposed solution reduces overhead by identifying the portion of threads that are unreliable and by applying any detection/protection mechanism only on them instead of the entire kernel. In other words, our solution reduces overhead by identifying the portion of threads that are unreliable and by organizing them into warps that consist of threads that are either reliable or unreliable. Then, any detection/protection technique (including RMT and TMR) can be applied upon the identified unreliable warps only, instead of the entire kernel.
Even with the simplest error detection and correction technique (warp duplication and triplication), we achieve an average 
saving of 20.61\% and 27.15\% execution cycles for error detection and error correction, respectively.


\section*{Data Availability}
This paper is based on already available open-source benchmark data and existing fault injection tools. The proposed technique offers a methodology for re-organizing threads after evaluating thread resilience using the fault site pruning methodology ~\cite{nie2018fault}. Any resilience evaluation technique in the literature can be also used to derive thread resilience~\cite{hari2015sassifi,nvbitfi,tselonis2016gufi,fang2014gpu} to guide remapping.

\section*{Acknowledgments}
We thank the anonymous reviewers for their insightful comments.
This material is based upon work supported by the National
Science Foundation (NSF) grant (\#1717532). This work
was performed in part using computing facilities at William \& Mary which were provided by contributions from NSF, the Commonwealth of Virginia Equipment
Trust Fund, and the Office of Naval Research.

\bibliographystyle{ieeetr}
\bibliography{reference}

\begin{thebibliography}{10}

\bibitem{DBLP:conf/dsn/FratinOLSRR18}
V.~Fratin, D.~A.~G. de~Oliveira, C.~B. Lunardi, F.~Santos, G.~Rodrigues, and
  P.~Rech, ``Code-dependent and architecture-dependent reliability behaviors,''
  in {\em {DSN}}, pp.~13--26, 2018.

\bibitem{killi}
S.~Ganapathy, J.~Kalamatianos, B.~M. Beckmann, S.~Raasch, and L.~G. Szafaryn,
  ``Killi: Runtime fault classification to deploy low voltage caches without
  {MBIST},'' in {\em 25th {IEEE} International Symposium on High Performance
  Computer Architecture, {HPCA} 2019, Washington, DC, USA, February 16-20,
  2019}, pp.~304--316, {IEEE}, 2019.

\bibitem{eklund2013medical}
A.~Eklund, P.~Dufort, D.~Forsberg, and S.~M. LaConte, ``Medical image
  processing on the {GPU}--past, present and future,'' {\em Medical image
  analysis}, vol.~17, no.~8, pp.~1073--1094, 2013.

\bibitem{pratx2011gpu}
G.~Pratx and L.~Xing, ``{GPU} computing in medical physics: A review,'' {\em
  Medical physics}, vol.~38, no.~5, pp.~2685--2697, 2011.

\bibitem{wenmei-mri}
S.~S. Stone, J.~P. Haldar, S.~C. Tsao, W.~mei W.~Hwu, B.~P. Sutton, and Z.-P.
  Liang, ``{Accelerating advanced MRI reconstructions on GPUs},'' {\em J.
  Parallel Distrib. Comput.}, vol.~68, no.~10, pp.~1307--1318, 2008.

\bibitem{govet-health}
R.~Foster, ``How to harness big data for improving public health,'' {\em
  Government Health IT}, 2012.

\bibitem{schmerken2009wall}
I.~Schmerken, ``Wall street accelerates options analysis with {GPU}
  technology,'' {\em Wall Street Technology}, vol.~11, 2009.

\bibitem{compfin}
NVIDIA, ``Computational finance.''

\bibitem{gpuneural}
NVIDIA, ``Researchers deploy {GPUs} to build world's largest artificial neural
  network.''

\bibitem{park2008low}
J.-H. Park, M.~Tada, D.~Kuzum, P.~Kapur, H.-Y. Yu, K.~C. Saraswat, {\em
  et~al.}, ``Low temperature (≤ 380° c) and high performance ge cmos
  technology with novel source/drain by metal-induced dopants activation and
  high-k/metal gate stack for monolithic 3d integration,'' in {\em Electron
  Devices Meeting, 2008. IEDM 2008. IEEE International}, pp.~1--4, IEEE, 2008.

\bibitem{nie2016large}
B.~Nie, D.~Tiwari, S.~Gupta, E.~Smirni, and J.~H. Rogers, ``A large-scale study
  of soft-errors on gpus in the field,'' in {\em High Performance Computer
  Architecture (HPCA), 2016 IEEE International Symposium on}, pp.~519--530,
  IEEE, 2016.

\bibitem{NieMASCOTS2017}
B.~Nie, J.~Xue, S.~Gupta, C.~Engelmann, E.~Smirni, and D.~Tiwari,
  ``Characterizing temperature, power, and soft-error behaviors in data center
  systems: Insights, challenges, and opportunities,'' in {\em 25th {IEEE}
  International Symposium on Modeling, Analysis, and Simulation of Computer and
  Telecommunication Systems, {MASCOTS} 2017, Banff, AB, Canada, September
  20-22, 2017}, pp.~22--31, 2017.

\bibitem{NieDSN18}
B.~Nie, J.~Xue, S.~Gupta, T.~Patel, C.~Engelmann, E.~Smirni, and D.~Tiwari,
  ``Machine learning models for {GPU} error prediction in a large scale {HPC}
  system,'' in {\em 48th Annual {IEEE/IFIP} International Conference on
  Dependable Systems and Networks, {DSN} 2018, Luxembourg City, Luxembourg,
  June 25-28, 2018}, pp.~95--106, 2018.

\bibitem{NieJS20}
B.~Nie, A.~Jog, and E.~Smirni, ``Characterizing accuracy-aware resilience of
  {GPGPU} applications,'' in {\em 20th {IEEE/ACM} International Symposium on
  Cluster, Cloud and Internet Computing, {CCGRID} 2020, Melbourne, Australia,
  May 11-14, 2020}, pp.~111--120, {IEEE}, 2020.

\bibitem{nvidia2009fermi}
``{NVIDIA Fermi Architecture Whitepaper}.''

\bibitem{nvidia2014kepler}
``{NVIDIA Kepler GK110 Architecture Whitepaper}.''

\bibitem{nvidia2016pascal}
``{GP100 Pascal Whitepaper}.''

\bibitem{hari2015sassifi}
S.~K.~S. Hari, T.~Tsai, M.~Stephenson, S.~W. Keckler, and J.~Emer, ``{SASSIFI}:
  Evaluating resilience of {GPU} applications,'' in {\em Proceedings of the
  Workshop on Silicon Errors in Logic-System Effects}, 2015.

\bibitem{takizawa2009checuda}
H.~Takizawa, K.~Sato, K.~Komatsu, and H.~Kobayashi, ``Checuda: A
  checkpoint/restart tool for cuda applications,'' in {\em 2009 International
  Conference on Parallel and Distributed Computing, Applications and
  Technologies}, pp.~408--413, IEEE, 2009.

\bibitem{laosooksathit2010lightweight}
S.~Laosooksathit, N.~Naksinehaboon, C.~Leangsuksan, A.~Dhungana, C.~Chandler,
  K.~Chanchio, and A.~Farbin, ``Lightweight checkpoint mechanism and modeling
  in gpgpu environment,'' {\em Computing (HPC Syst)}, vol.~12, no.~2010, 2010.

\bibitem{wadden2014real}
J.~Wadden, A.~Lyashevsky, S.~Gurumurthi, V.~Sridharan, and K.~Skadron,
  ``Real-world design and evaluation of compiler-managed {GPU} redundant
  multithreading,'' {\em ACM SIGARCH Computer Architecture News}, vol.~42,
  no.~3, pp.~73--84, 2014.

\bibitem{gupta2017compiler}
M.~Gupta, D.~Lowell, J.~Kalamatianos, S.~Raasch, V.~Sridharan, D.~Tullsen, and
  R.~Gupta, ``Compiler techniques to reduce the synchronization overhead of gpu
  redundant multithreading,'' in {\em 2017 54th ACM/EDAC/IEEE Design Automation
  Conference (DAC)}, pp.~1--6, IEEE, 2017.

\bibitem{mahmoud2018optimizing}
A.~Mahmoud, S.~K.~S. Hari, M.~B. Sullivan, T.~Tsai, and S.~W. Keckler,
  ``Optimizing software-directed instruction replication for gpu error
  detection,'' in {\em SC18: International Conference for High Performance
  Computing, Networking, Storage and Analysis}, pp.~842--853, IEEE, 2018.

\bibitem{nie2018fault}
B.~Nie, L.~Yang, A.~Jog, and E.~Smirni, ``Fault site pruning for practical
  reliability analysis of gpgpu applications,'' in {\em 2018 51st Annual
  IEEE/ACM International Symposium on Microarchitecture (MICRO)}, pp.~749--761,
  IEEE, 2018.

\bibitem{LishanSigmetrcis2021}
L.~Yang, B.~Nie, A.~Jog, and E.~Smirni, ``Sugar: Speeding up gpgpu application
  resilience estimation with input sizing,'' {\em Proc. {ACM} Meas. Anal.
  Comput. Syst.}, vol.~5, no.~1, pp.~1:1--1:29, 2021.

\bibitem{DBLP:conf/dsn/LiP18}
G.~Li and K.~Pattabiraman, ``Modeling input-dependent error propagation in
  programs,'' in {\em 48th Annual {IEEE/IFIP} International Conference on
  Dependable Systems and Networks, {DSN} 2018, Luxembourg City, Luxembourg,
  June 25-28, 2018}, pp.~279--290, {IEEE} Computer Society, 2018.

\bibitem{yazdanbakhsh2016axbench}
A.~Yazdanbakhsh, D.~Mahajan, H.~Esmaeilzadeh, and P.~Lotfi-Kamran, ``Axbench: A
  multiplatform benchmark suite for approximate computing,'' {\em IEEE Design
  \& Test}, vol.~34, no.~2, pp.~60--68, 2016.

\bibitem{cudagdb}
``{CUDA-GDB}.''

\bibitem{grauer2012auto}
S.~Grauer-Gray, L.~Xu, R.~Searles, S.~Ayalasomayajula, and J.~Cavazos,
  ``Auto-tuning a high-level language targeted to gpu codes,'' in {\em
  Innovative Parallel Computing (InPar), 2012}, pp.~1--10, IEEE, 2012.

\bibitem{che2009rodinia}
S.~Che, M.~Boyer, J.~Meng, D.~Tarjan, J.~W. Sheaffer, S.-H. Lee, and
  K.~Skadron, ``Rodinia: A benchmark suite for heterogeneous computing,'' in
  {\em 2009 IEEE International Symposium on Workload Characterization (IISWC)},
  pp.~44--54, Ieee, 2009.

\bibitem{book}
D.~B. Kirk and W.~H. Wen-Mei, {\em Programming massively parallel processors: a
  hands-on approach}.
\newblock Morgan kaufmann, 2016.

\bibitem{owl-asplos13}
A.~Jog, O.~Kayiran, N.~Chidambaram~Nachiappan, A.~K. Mishra, M.~T. Kandemir,
  O.~Mutlu, R.~Iyer, and C.~R. Das, ``{OWL}: cooperative thread array aware
  scheduling techniques for improving {GPGPU} performance,'' in {\em ACM
  SIGPLAN Notices}, vol.~48, pp.~395--406, ACM, 2013.

\bibitem{fang2014gpu}
B.~Fang, K.~Pattabiraman, M.~Ripeanu, and S.~Gurumurthi, ``{GPU-Qin}: A
  methodology for evaluating the error resilience of {GPGPU} applications,'' in
  {\em Performance Analysis of Systems and Software (ISPASS), 2014 IEEE
  International Symposium on}, pp.~221--230, IEEE, 2014.

\bibitem{llfi-gpu}
G.~Li, K.~Pattabiraman, C.-Y. Cher, and P.~Bose, ``Understanding error
  propagation in {GPGPU} applications,'' in {\em High Performance Computing,
  Networking, Storage and Analysis, SC16: International Conference for},
  pp.~240--251, IEEE, 2016.

\bibitem{DBLP:conf/dsn/SangchooliePK17}
B.~Sangchoolie, K.~Pattabiraman, and J.~Karlsson, ``One bit is (not) enough: An
  empirical study of the impact of single and multiple bit-flip errors,'' in
  {\em 47th Annual {IEEE/IFIP} International Conference on Dependable Systems
  and Networks, {DSN} 2017, Denver, CO, USA, June 26-29, 2017}, pp.~97--108,
  {IEEE} Computer Society, 2017.

\bibitem{tselonis2016gufi}
S.~Tselonis and D.~Gizopoulos, ``Gufi: A framework for gpus reliability
  assessment,'' in {\em Performance Analysis of Systems and Software (ISPASS),
  2016 IEEE International Symposium on}, pp.~90--100, IEEE, 2016.

\bibitem{gpgpu-sim}
A.~Bakhoda, G.~L. Yuan, W.~W. Fung, H.~Wong, and T.~M. Aamodt, ``Analyzing
  {CUDA} workloads using a detailed {GPU} simulator,'' in {\em Performance
  Analysis of Systems and Software, 2009. ISPASS 2009. IEEE International
  Symposium on}, pp.~163--174, IEEE, 2009.

\bibitem{nvbitfi}
``Nvbitfi.'' \url{https://github.com/NVlabs/nvbitfi}.

\bibitem{yang2020practical}
L.~Yang, B.~Nie, A.~Jog, and E.~Smirni, ``Practical resilience analysis of
  gpgpu applications in the presence of single-and multi-bit faults,'' {\em
  IEEE Transactions on Computers}, vol.~70, no.~1, pp.~30--44, 2021.

\bibitem{vallero2019combining}
A.~Vallero and S.~Di~Carlo, ``Combining cluster sampling and ace analysis to
  improve fault-injection based reliability evaluation of gpu-based systems,''
  in {\em 2019 IEEE International Symposium on Defect and Fault Tolerance in
  VLSI and Nanotechnology Systems (DFT)}, pp.~8138--8143, IEEE, 2019.

\bibitem{li2018modeling}
G.~Li and K.~Pattabiraman, ``Modeling input-dependent error propagation in
  programs,'' in {\em 2018 48th Annual IEEE/IFIP International Conference on
  Dependable Systems and Networks (DSN)}, pp.~279--290, IEEE, 2018.

\bibitem{yusuf2015chiminey}
I.~I. Yusuf, I.~E. Thomas, M.~Spichkova, S.~Androulakis, G.~R. Meyer, D.~W.
  Drumm, G.~Opletal, S.~P. Russo, A.~M. Buckle, and H.~W. Schmidt, ``Chiminey:
  Reliable computing and data management platform in the cloud,'' in {\em 2015
  IEEE/ACM 37th IEEE International Conference on Software Engineering}, vol.~2,
  pp.~677--680, IEEE, 2015.

\bibitem{de2020delta}
J.~De~Bleser, D.~Di~Nucci, and C.~De~Roover, ``A delta-debugging approach to
  assessing the resilience of actor programs through run-time test
  perturbations,'' in {\em Proceedings of the IEEE/ACM 1st International
  Conference on Automation of Software Test}, pp.~21--30, 2020.

\bibitem{chan2017ipa}
A.~Chan, S.~Winter, H.~Saissi, K.~Pattabiraman, and N.~Suri, ``Ipa: Error
  propagation analysis of multi-threaded programs using likely invariants,'' in
  {\em 2017 IEEE International Conference on Software Testing, Verification and
  Validation (ICST)}, pp.~184--195, IEEE, 2017.

\bibitem{yang2020far}
Y.~Yang, Y.~Wu, K.~Pattabiraman, L.~Wang, and Y.~Li, ``How far have we come in
  detecting anomalies in distributed systems? an empirical study with a
  statement-level fault injection method,'' in {\em 2020 IEEE 31st
  International Symposium on Software Reliability Engineering (ISSRE)},
  pp.~59--69, IEEE, 2020.

\bibitem{dimitrov2009understanding}
M.~Dimitrov, M.~Mantor, and H.~Zhou, ``Understanding software approaches for
  gpgpu reliability,'' in {\em Proceedings of 2nd Workshop on General Purpose
  Processing on Graphics Processing Units}, pp.~94--104, ACM, 2009.

\bibitem{milluzzi2017exploration}
A.~Milluzzi and A.~George, ``Exploration of {TMR} fault masking with persistent
  threads on tegra gpu socs,'' in {\em 2017 IEEE Aerospace Conference},
  pp.~1--7, IEEE, 2017.

\bibitem{chen2016gpu}
J.~Chen, S.~Li, and Z.~Chen, ``{GPU-ABFT}: Optimizing algorithm-based fault
  tolerance for heterogeneous systems with {GPUs},'' in {\em 2016 IEEE
  International Conference on Networking, Architecture and Storage (NAS)},
  pp.~1--2, IEEE, 2016.

\end{thebibliography}

\end{document}